\documentclass[12pt]{article}

\usepackage[dvipsname]{xcolor}

\usepackage{graphicx}
\usepackage{amsmath}        
\usepackage{amssymb}        
\usepackage{slashed}
\usepackage{bm}

\def\mydate{Published in PRD 104, 115018 (2021)}
\def\ignore#1{{}}

\def\go{\rightarrow}
\def\dd{\partial}

\def\ep{{\epsilon}}

\def\eff{{\rm eff}}
\def\SM{{\rm SM}}
\def\KK{{\rm KK}}
\def\EM{{\rm EM}}
\def\EW{{\rm EW}}

\def\LR{{\rm LR}}
\def\decay{{\rm decay}}
\def\min{{\rm min}}
\def\max{{\rm max}}
\def\onehalf{\hbox{$\frac{1}{2}$}}
\def\la{\langle}
\def\ra{\rangle}

\def\E{\hbox{\bfseries \itshape E}}
\def\N{\hbox{\bfseries \itshape N}}

\def\mymat#1#2{\begin{matrix}#1 \cr \noalign{\kern -2pt} #2\end{matrix}}

\def\mynoalign{\noalign{\kern 4pt}}
\def\mysnoalign{\noalign{\kern 3pt}}

\makeatletter
\@addtoreset{equation}{section}
\makeatother

\begin{document}

\thispagestyle{empty}

{\small \noindent \mydate    \hfill OU-HET-1086}

{\small \noindent    \hfill KYUSHU-HET-224}

\vskip 2.cm

\baselineskip=30pt plus 1pt minus 1pt

\begin{center}
{\Large \bf  Electroweak and Left-Right Phase Transitions}

{\Large \bf   in  $\bm{SO(5)\times U(1) \times SU(3)}$ Gauge-Higgs Unification}

\end{center}


\baselineskip=22pt plus 1pt minus 1pt

\vskip 1.cm

\begin{center}
{\bf Shuichiro Funatsu$^1$, Hisaki Hatanaka$^2$,  Yutaka Hosotani$^3$,}

{\bf Yuta Orikasa$^4$ and Naoki Yamatsu$^5$}

\baselineskip=18pt plus 1pt minus 1pt

\vskip 10pt
{\small \it $^1$Institute of Particle Physics and Key Laboratory of Quark and Lepton 
Physics (MOE), Central China Normal University, Wuhan, Hubei 430079, China} \\
{\small \it $^2$Osaka, Osaka 536-0014, Japan} \\
{\small \it $^3$Department of Physics, Osaka University, 
Toyonaka, Osaka 560-0043, Japan} \\
{\small \it $^4$Institute of Experimental and Applied Physics, Czech Technical University in Prague,} \\
{\small \it Husova 240/5, 110 00 Prague 1, Czech Republic} \\
{\small \it $^5$Department of Physics, Kyushu University, Fukuoka 819-0395, Japan} \\

\end{center}

\vskip 1.cm
\baselineskip=18pt plus 1pt minus 1pt

\begin{abstract}
The electroweak phase transition in  GUT inspired $SO(5)\times U(1) \times SU(3)$ 
gauge-Higgs unification  is shown to be  weakly first order and occurs at 
$T = T_c^{\rm EW} \sim 163\,$GeV, which is very similar to the behavior in the
standard model in perturbation theory.  A new phase appears at higher temperatures.  
$SU(2)_L \times U(1)_Y$ ($\theta_H=0$) and $SU(2)_R \times U(1)_{Y'}$ ($\theta_H=\pi$) phases 
become almost degenerate above $T \sim m_\KK$ where $m_\KK$ is the Kaluza-Klein mass scale 
typically around 13$\,$TeV and $\theta_H$ is the Aharonov-Bohm phase along the fifth dimension.  
The two phases become degenerate at $T = T_c^{\rm LR} \sim m_\KK$.   As the temperature 
drops in the evolution of the early Universe the $SU(2)_R \times U(1)_{Y'}$ phase becomes unstable.
The tunneling rate from the $SU(2)_R \times U(1)_{Y'}$ phase  to the $SU(2)_L \times U(1)_Y$ phase 
becomes sizable and a first-order phase transition takes place at $T=2.5 \sim 2.6\,$TeV.
The amount of gravitational waves produced in this left-right phase transition is small, far below 
the reach of the sensitivity of Laser Interferometer Space Antenna (LISA).
A detailed analysis of the $SU(2)_R \times U(1)_{Y'}$ phase is also given.
It is shown that the $W$ boson, $Z$ boson and photon, with $\theta_H$ varying from 0 to $\pi$, 
are transformed to gauge bosons in the $SU(2)_R \times U(1)_{Y'}$ phase.
Gauge couplings and wave functions of quarks, leptons  and dark fermions in the 
$SU(2)_R \times U(1)_{Y'}$ phase are determined.
\end{abstract}


\newpage

\baselineskip=20pt plus 1pt minus 1pt
\parskip=0pt

\section{Introduction} 

The standard model (SM), $SU(3)_C \times SU(2)_L \times U(1)_Y$ gauge theory, has been 
firmly established at low energies.  Implications of the model in the history of the Universe
have also been discussed intensively.  
There remain a few important
mysteries such as dark matter and baryon number generation in the Universe.
Something beyond the SM is necessary.

The Higgs boson with a mass of 125$\,$GeV was discovered in 2012, whose properties (observed so far) 
have been consistent with the SM within experimental errors.  Yet it is not clear whether  the observed
boson is precisely what the SM assumes to exist.  We need further measurements of the Higgs boson's couplings
to itself and other particles to make a judgment.  With the possibility of the appearance of  cracks of the SM
in mind, many alternative models with an extension of the  scalar sector have been proposed.

Furthermore the Higgs sector in the SM has the severe gauge hierarchy problem when implemented
in a larger theory such as grand unification.
One possible answer to this problem is  gauge-Higgs unification  (GHU) in which the Higgs boson is 
identified with a zero mode of the fifth dimensional component of the gauge potential. \cite{Hosotani1983}-\cite{Scrucca2003}
It appears as a fluctuation mode of an Aharonov-Bohm  (AB) phase $\theta_H$  in  the fifth dimension.
The  mass of the Higgs boson is generated at the one-loop level, the finiteness of which is
guaranteed by gauge invariance.   Among various GHU models $SO(5) \times U(1)_X \times SU(3)_C$ 
GHU model in the Randall-Sundrum (RS) warped space reproduces nearly the same phenomenology 
at low energies as the SM for $\theta_H \lesssim 0.1$.\cite{ACP2005}-\cite{FHHOS2013}
Gauge couplings of quarks and leptons are almost the same as in the SM.  
The Cabibbo-Kobayashi-Maskawa (CKM) mass mixing is
incorporated with natural suppression of flavor-changing neutral currents  (FCNCs).   
Yukawa couplings  of quarks and leptons are suppressed
by a factor of $\cos \theta_H$ or $\cos^2 \onehalf \theta_H$ compared to those in the SM.  
The model predicts $Z'$ bosons as Kaluza-Klein (KK) excited modes of $\gamma$, $Z$ and $Z_R$ where
$Z_R$ is the gauge boson of $SU(2)_R$ in $SO(4) \simeq SU(2)_L \times SU(2)_R \subset SO(5)$.
It has been shown that effects of  $Z'$ bosons can be   observed in fermion pair production at electron-positron ($e^- e^+$) 
collider experiments.  Significant interference effects should be seen even in the early stage of the
planned International Linear Collider (ILC) experiments at energies of 250$\,$GeV by measuring the dependence 
on polarization of  electron and positron beams.\cite{GUTinspired2019a}-\cite{Funatsu2019a}

Natural questions arise about the behavior of  $SO(5) \times U(1)_X \times SU(3)_C$ GHU at finite temperature.
At which temperature is the electroweak (EW) symmetry, $SU(2)_L \times U(1)_Y$, restored?  Is the transition first order
or second order?  Is there any difference from the SM?\cite{Quiros1999}-\cite{Onofrio2016}
These are main themes analyzed in this paper.
Previously phase transitions in GHU have been analyzed in various models.\cite{Adachi2020}-\cite{Hatanaka2013}
Adachi and Maru analyzed
their $SU(3) \times U(1)$ GHU model to show that the EW phase transition is
strongly first order [$v_c/T_c = O(1)$] though $v_c$ and $T_c$  turn out about $1\,$GeV,  being too small.
In  $SO(5) \times U(1)_X \times SU(3)_C$ GHU the $SU(2)_L \times U(1)_Y$ symmetric phase corresponds
to $\theta_H = 0$, which dynamically breaks down to $U(1)_\EM$ with nonvanishing $\theta_H \sim 0.1$
at zero temperature.  It will be shown below that  the $SU(2)_L \times U(1)_Y$
symmetry is restored around $T = T_c^\EW \sim 163\,$GeV.    The transition is shown to be  weakly first order,
just as in the SM in perturbation theory.
A new phase emerges at higher temperatures.  In $SO(5) \times U(1)_X \times SU(3)_C$ GHU, $SO(5)$ symmetry
breaks down to  $SO(4) \simeq SU(2)_L  \times SU(2)_R$ by orbifold boundary conditions.
Although a brane scalar at the UV brane spontaneously breaks $SU(2)_R \times U(1)_X$ to $U(1)_Y$
at $T=0$, a new minimum of the effective potential at finite temperature appears at $\theta_H = \pi$.
It will be seen that above $T=T_{c1}^\LR \sim m_\KK$ the $\theta_H=0$ phase and the $\theta_H=\pi$ phase
become almost degenerate.  The two phases are separated by a barrier so that a domain structure will be formed
in the Universe  as the Universe expands and the temperature drops to around $T_{c1}^\LR$.
As the Universe cools down further  the $\theta_H=\pi$ phase becomes absolutely unstable 
at $T=T_{c2}^\LR \sim 2.3\,$TeV.
The transition from the $\theta_H=\pi$  phase to the $\theta_H=0$  phase takes place by tunneling 
at $T=T_{\rm decay}^\LR \sim 2.5\,$TeV.
In the $\theta_H=0$  phase there is $SU(2)_L \times U(1)_Y$ gauge invariance, while in 
the $\theta_H=\pi$  phase  $SU(2)_R \times U(1)_{Y'}$ gauge symmetry emerges.
The transition from the $\theta_H=\pi$ phase to the $\theta_H=0$ phase is called 
as the left-right (LR)  transition.

In this paper  we consider GHU models defined in the RS warped space
with fixed curvature and size.\cite{RS1}    The stabilization of  the RS warped space can be achieved 
by the Goldberger-Wise mechanism.\cite{Goldberger1999}  
It has been discussed in the literature  that  in this case a  decompactification phase transition 
from the RS space to the black-brane phase 
may take place below the KK scale.\cite{Creminelli2002, Agashe2020}
The discussion of the LR transition  in this paper remains valid  as long as the RS warped space is stable 
around  $T=T_{\rm decay}^\LR$.

The paper is organized as follows.   In Section 2 the $SO(5) \times U(1)_X \times SU(3)_C$ GHU model is 
introduced and its effective potential $V_\eff (\theta_H; T)$ at finite temperature is given in Section 3.
The EW phase transition is examined in Section 4,  and  the LR transition is studied in Section 5. 
The LR transition is the transition from the $\theta_H=\pi$  phase to the $\theta_H=0$ phase.
We give a detailed analysis of the $\theta_H=\pi$  phase in Section 6.  It will be shown 
that the $\theta_H=\pi$ phase corresponds to  $SU(2)_R \times U(1)_{Y'}$ gauge symmetry, 
which becomes manifest in the twisted gauge.  
Gauge couplings of quarks, leptons, and dark fermions at $\theta_H = 0$ and $\theta_H =\pi$ are clarified.
Wave functions of those fields are summarized in Appendixes.
Section 7 is devoted to a summary and discussion.

\section{Model} 

We analyze  $SO(5) \times U(1)_X \times SU(3)_C$ GHU models defined in the RS
warped space. We focus on the GUT inspired $SO(5) \times U(1)_X \times SU(3)_C$ GHU model
specified in Refs.\ \cite{GUTinspired2019a, FCNC2020a, GUTinspired2020b}.
The metric of the RS space is given by
\begin{align}
ds^2=  \frac{1}{z^2} \bigg(\eta_{\mu\nu}dx^{\mu} dx^{\nu} + \frac{dz^2}{k^2}\bigg) 
\quad {\rm for~} 1 \le z \le z_L ~.
\label{RS-metric}
\end{align}
The bulk region $1<z<z_L$  is anti-de Sitter (AdS) spacetime 
with a cosmological constant $\Lambda=-6k^2$, which is sandwiched by the
UV brane at $z=1$  and the IR brane at $z=z_L$.  
The KK mass scale is $m_{\rm KK}=\pi k/(z_L-1) \simeq \pi kz_L^{-1}$ for $z_L\gg 1$.
In addition to the gauge fields $A_M^{SU(3)_C}$, $A_M^{SO(5)}$, and $A_M^{U(1)_X}$ of 
$SU(3)_C$,  $SO(5)$ and $U(1)_X$, matter fields are introduced in the bulk and on the UV brane.
They are summarized in Table \ref{Tab:matterlist}.

\begin{table}[tbh]
\renewcommand{\arraystretch}{1.2}
\begin{center}
\begin{tabular}{|c|c|c|c|c|c|c|}
\hline
\multicolumn{2}{|c|}{Field} & ${\cal G}$&$G_{22}$ &Left &Right &Name\\
\hline
quark &$\Psi_{({\bf 3,4})}^{\alpha}$ &$({\bf 3,4})_{\frac{1}{6}}$&$[{\bf 2,1}]$
&$(+,+)$ &$(-,-)$ &$\begin{matrix} u & c & t \cr d & s & b\end{matrix}$\\
\cline{4-7}
&&&$[{\bf 1,2}]$
&$(-,-)$ &$(+,+)$ &$\begin{matrix} u'  & c' & t' \cr d' & s' & b' \end{matrix}$\\
\cline{2-7}
&$\Psi_{({\bf 3,1})}^{\pm \alpha}$ &$({\bf 3,1})_{-\frac{1}{3}}$&$[{\bf 1,1}]$
&$(\pm ,\pm )$ &$(\mp , \mp )$ &$D^{\pm}_d ~ D^{\pm}_s ~ D^{\pm}_b$\\
\hline
lepton &$\Psi_{({\bf 1,4})}^{\alpha}$ &$({\bf 1,4})_{-\frac{1}{2}}$&$[{\bf 2,1}]$ 
&$(+,+)$ &$(-,-)$ &$\begin{matrix} \nu_e  & \nu_\mu & \nu_\tau \cr e & \mu & \tau \end{matrix}$\\
\cline{4-7}
&&&$[{\bf 1,2}]$
&$(-,-)$ &$(+,+)$ &$\begin{matrix} \nu_e'  & \nu_\mu' & \nu_\tau' \cr e' & \mu' & \tau' \end{matrix}$\\
\hline
darkF &$\Psi_F^\beta$ &$({\bf 3,4})_{\frac{1}{6}}$ &$[{\bf 2,1}]$  &$(-,+)$ &$(+,-)$ &$F$ \\
\cline{4-7}
&&&$[{\bf 1,2}]$ &$(+,-)$ &$(-,+)$ &$F'$ \\
\hline
darkF$_\ell$ &$\Psi_{F_\ell}^\beta$ &$({\bf 1,4})_{-\frac{1}{2}}$ &$[{\bf 2,1}]$  &$(-,+)$ &$(+,-)$ &$F_\ell$ \\
\cline{4-7}
&&&$[{\bf 1,2}]$ &$(+,-)$ &$(-,+)$ &$F_\ell'$ \\
\hline
darkV &$\Psi_V^{\pm \gamma}$ &$({\bf 1}, {\bf 5})_{0}$ &$[{\bf 2,2}]$ &$(\pm,\pm)$ &$(\mp,\mp)$ 
&$\begin{matrix}N^\pm &  E^{\prime \pm} \cr E^\pm &  N^{\prime \pm} \end{matrix}$ \\
\cline{4-7}
&&&$[{\bf 1,1}]$ &$(\mp,\mp)$ &$(\pm,\pm)$ &$S^\pm$ \\
\hline
brane fermion &$\chi^\alpha$ &$({\bf 1}, {\bf 1})_{0}$ &$[{\bf 1,1}]$ &$\cdots$ &$\cdots$ &$\chi$ \\
\hline
brane scalar &$\Phi_S$ &$ ({\bf 1}, {\bf 4})_{\frac{1}{2}}$ &$[{\bf 2,1}]$  &$\cdots$ &$\cdots$ &$\Phi_{[2,1]}$ \\
\cline{4-7}
&&&$[{\bf 1,2}]$ &$\cdots$ &$\cdots$ &$\Phi_{[1,2]}$ \\
\hline
\end{tabular}
\caption{\small
Matter fields in the bulk and on the UV brane.
Content of each field in 
${\cal G} = SU(3)_C\times SO(5) \times U(1)_X$ and
$G_{22} = SU(2)_L\times SU(2)_R(\subset SO(5))$  is shown.
Parity assignment $(P_0, P_1)$ of left- and right-handed quarks, leptons and dark fermion 
multiplets in the bulk is shown. 
}
\label{Tab:matterlist}
\end{center}
\end{table}

The action of the model is given in Refs.\ \cite{GUTinspired2019a, GUTinspired2020b}.
It has been shown that the model reproduces the SM phenomenology at low energies.  
The bulk part of the action for the fermion multiplets are given, 
with  $\overline{\Psi} = i \Psi^\dagger \gamma^0$, by 
\begin{align}
&S_{\rm bulk}^{\rm fermion} =  \int d^5x\sqrt{-\det G} \, \bigg\{ 
\sum_J  \overline{\Psi}{}^J   {\cal D} (c_J) \Psi^J    \cr
\noalign{\kern 5pt}
&\hskip  1.0cm
-  \sum_\alpha \Big( m_{D_\alpha} \overline{\Psi}{}_{\bf (3,1)}^{+ \alpha} \Psi_{\bf (3,1)}^{- \alpha}  + {\rm H.c.} \Big)
-  \sum_\gamma \Big( m_{V_\gamma} \overline{\Psi}{}_{\bf (1,5)}^{+ \gamma} \Psi_{\bf (1,5)}^{- \gamma}  + {\rm H.c.} \Big)
 \bigg\} , \cr
\noalign{\kern 5pt}
&{\cal D}(c)= \gamma^A {e_A}^M
\bigg( D_M+\frac{1}{8}\omega_{MBC}[\gamma^B,\gamma^C]  \bigg) - c \, k ~, \cr
\noalign{\kern 5pt}
&D_M =  \dd_M - ig_S A_M^{SU(3)} -i g_A A_M^{SO(5)}  -i g_B Q_X A_M ^{U(1)} ~, 
\label{fermionAction1}
\end{align} 
where the sum $\sum_J$ extends over $ \Psi^J = \Psi_{\bf (3,4)}^\alpha$, $ \Psi_{\bf (1,4)}^\alpha$,
$\Psi_{\bf (3,1)}^{\pm \alpha}$, $ \Psi_F^\beta$, $ \Psi_{F_\ell}^\beta$ and $ \Psi_V^{\pm \gamma}$. 
The bulk mass parameter $c_J$ of each fermion multiplet is important to specify a mass and wave function 
of the lowest (zero) mode.   In the GUT inspired model  bulk mass parameters of $ \Psi_{\bf (3,4)}^\alpha$ and
$ \Psi_{\bf (1,4)}^\alpha$ are taken to be negative.
$\Psi_{({\bf 3,1})}^{\pm \alpha}$ and $ \Psi_V^{\pm \gamma}$
have additional Dirac-type masses $m_{D_\alpha}$ and $m_{V_\gamma}$, respectively.

The original RS metric  is  given by
\begin{align}
ds^2= e^{-2\sigma(y)} \eta_{\mu\nu}dx^\mu dx^\nu+dy^2
\label{RSmetric2}
\end{align}
where $\eta_{\mu\nu}=\mbox{diag}(-1,+1,+1,+1)$, $\sigma(y)=\sigma(y+ 2L)=\sigma(-y)$ 
and $\sigma(y)=ky$ for $0 \le y \le L$.
The coordinates $y$ and $z$  are related  by $z = e^{ky}$ for $0 \le y \le L$.
The fifth dimension in the RS space has topology of $S^1/Z_2$.  
In the $y$ coordinate the orbifold boundary conditions are given by
$(A_\mu, A_y)(x, y_j -y) = P_j (A_\mu,  - A_y)(x, y_j + y)  P_j$ ($j=0,1$) where $(y_0, y_1) = (0, L)$
and $(P_j)^2 = 1$.  It follows that $A_y (x, y+ 2L) = P_1 P_0 A_y (x, y) P_0 P_1$.
We take the orbifold boundary conditions $P_0 = P_1 = \mbox{diag}(1,1,1,1,-1)$, which  breaks 
$SO(5)$ to $SO(4) \simeq SU(2)_L \times SU(2)_R$.
Further the brane scalar field $\Phi_S$ located at $z=1$ develops a nonvanishing expectation value
to spontaneously break $SU(2)_R \times U(1)_X$ to $U(1)_Y$.

The 4D Higgs boson $\Phi_H (x)$ is the zero mode of the $SO(5)/SO(4)$ part of $A_z^{SO(5)}$,
\begin{align}
A_z^{(j5)} (x, z) &= \frac{1}{\sqrt{k}} \, \phi_j (x) u_H (z) + \cdots  \quad (j= 1 \sim 4) , \cr
\noalign{\kern 5pt}
u_H (z) &= \sqrt{ \frac{2}{z_L^2 -1} } \, z ~,  \cr
\noalign{\kern 5pt}
\Phi_H (x) &= \frac{1}{\sqrt{2}} \begin{pmatrix} \phi_2 + i \phi_1 \cr \phi_4 - i\phi_3 \end{pmatrix} .
\label{4dHiggs}
\end{align}
At the quantum level $\Phi_H$ develops a nonvanishing expectation value.  Without loss of generality
we assume $\la \phi_1 \ra , \la \phi_2 \ra , \la \phi_3 \ra  =0$ and  $\la \phi_4 \ra \not= 0$, 
which is related to the Aharonov-Bohm (AB) phase $\theta_H$ in the fifth dimension.  
Eigenvalues of 
\begin{align}
\hat W &= P \exp \bigg\{   i g_A \int_{-L}^L dy \, A_y \bigg\}    \cr
\noalign{\kern 5pt}
&= P \exp \bigg\{ 2 i g_A \int_1^{z_L} dz \, A_z \bigg\}  
\label{ABphase1}
\end{align}
are gauge invariant.  
For $A_z = (2k)^{-1/2} \phi_4 (x) u_H (z) T^{(45)}$
\begin{align}
&\hat W = \exp \Big\{ i f_H^{-1} \phi_4 (x) \cdot 2 T^{(45)} \Big\} ~, \cr
\noalign{\kern 5pt}
&f_H = \frac{2}{g_A} \sqrt{ \frac{k}{z_L^2 -1}} = \frac{2}{g_w} \sqrt{ \frac{k}{L(z_L^2 -1)}} ~, \cr
\noalign{\kern 5pt}
&\theta_H = \frac{\la \phi_4 \ra}{f_H} ~,
\label{ABphase2}
\end{align}
where $g_w = g_A/\sqrt{L}$ is the 4D $SU(2)_L$ gauge coupling.

At zero temperature the effective potential  $V_\eff (\theta_H)$ has a global
minimum at $\theta_H \not= 0$ which breaks $SU(2)_L \times U(1)_Y$ to $U(1)_\EM$.
$W$ bosons, $Z$ bosons, quarks and leptons acquire masses with  $\theta_H \not= 0$.

The RS metric has two parameters, $k$ and $L$. With one of them (or the KK mass scale 
$m_\KK = \pi k (z_L -1)^{-1} \sim \pi k z_L^{-1}$) given, the other parameter is fixed by the $Z$ boson mass $m_Z$
once the resultant values of $\theta_H$ and the weak mixing angle $\sin^2 \theta_W$ are specified.  
The bulk mass parameters $c_J$ of quark multiplets $\Psi_{\bf (3,4)}^\alpha$ and lepton multiplets
$\Psi_{\bf (1,4)}^\alpha$ are determined from masses of up-type quarks and charged leptons.
Masses of down-type quarks are reproduced through bulk actions for 
$\Psi_{\bf (1,4)}^\alpha, \Psi_{\bf (3,1)}^{\pm \alpha}$ and brane interactions among
$\Psi_{\bf (1,4)}^\alpha$,  $\Psi_{\bf (3,1)}^{\pm \alpha}$, and $\Phi_S$.
It has been shown that the CKM  mixing  matrix can be generated
with natural suppression of FCNCs in the quark sector.\cite{FCNC2020a}
The brane fermions $\chi^\alpha$ are Majorana fermions.  Brane interactions among
$\chi^\alpha$, $\Psi_{\bf (1,4)}^\alpha$ and $\Phi_S$ induce gauge-Higgs seesaw mechanism\cite{seesaw2017}
similar to the inverse seesaw mechanism in grand unified theories.\cite{Mohapatra1986}  Tiny neutrino masses
are naturally explained.

Dark fermions are relevant to have dynamical electroweak symmetry breaking by the Hosotani
mechanism.   There are five parameters $(n_F, c_F, n_V, c_V, m_V)$ to be specified in the dark fermion sector 
where $n_F$ ($n_V$) and $c_F$ ($c_V$) are the number and bulk mass parameter of 
$\Psi_F$($\Psi_V^\pm$) multiplets, and $m_V$ is a Dirac-type mass in (\ref{fermionAction1}).
Rigorously speaking, there are additional parameters $(n_{F_\ell}, c_{F_\ell})$ associated with $\Psi_{F_\ell}$.
In the evaluation of the effective potential $V_\eff (\theta_H)$, contributions coming from $\Psi_{F_\ell}$
are summarized by the replacement $n_F \go n_F + \frac{1}{3} n_{F_\ell}$ for $c_{F_\ell} = c_F$.
For $|c_{F_\ell}|  > \onehalf$  its contributions are negligible.   As seen below, such physical quantities as
transition temperature do not depend on $n_F$ so much, and therefore we suppress the reference to 
$(n_{F_\ell}, c_{F_\ell})$ in discussing $V_\eff (\theta_H)$ below.
The parameters $(n_F, c_F, n_V, c_V, m_V)$ are chosen such that $V_\eff (\theta_H)$ has
a global minimum at a desired value of $\theta_H$ and the resultant Higgs boson mass 
$m_H = f_H^{-1} \{d^2  V_\eff (\theta) /d \theta^2 |_{\theta= \theta_H} \}^{1/2}$ is 125.1$\,$GeV.
This procedure leaves three parameters unfixed.
Surprisingly there appears the $\theta_H$ universality in physics at low energies.\cite{GUTinspired2020b}
Gauge couplings of quarks, leptons, and $W$ and $Z$ bosons are almost independent of $\theta_H$
and other parameters.  Yukawa couplings of quarks and leptons, Higgs couplings of $W$ and $Z$
are suppressed, compared to those in the SM, by a factor $\cos^2 \onehalf \theta_H$ or $\cos \theta_H$,
but do not depend on details of the parameters in the dark fermion sector.
Similarly cubic and quartic self-couplings of the Higgs boson become smaller than those in the SM,
depending solely on $\theta_H$, but not on the choice of the parameters in the dark fermion sector.
The resultant phenomenology at low energies is nearly the same as that in the SM.

Distinct signals of GHU appear in physics of KK excited modes of gauge fields and fermions.
For $\theta_H \sim 0.1$ the KK mass scale $m_\KK$ turns out in the range 10$\,$TeV to 15$\,$TeV. 
Masses of the first KK bosons of $\gamma$, $Z$, and $Z_R$, which play the role of $Z'$ bosons,  
are around $0.8 \, m_\KK$.  Couplings of quarks and leptons to those $Z'$ bosons exhibit large
parity violation, and couplings of either left-handed or right-handed quarks and leptons become
rather large.  It has been shown that a large deviation from the SM can be observed in the 
$e^- e^+ \go f \bar f$ processes, where $f$ is a lepton or quark, at energies  well below $m_{Z'}$. 
Significant deviation can be observed even in the early stage of the planned International Linear
Collider (ILC) at $\sqrt{s} = 250\,$GeV.  The interference effect between the two amplitudes for
$e^- e^+ \go \gamma, Z \go  f \bar f$ and $e^- e^+ \go Z' \go  f \bar f$ becomes very large,
and cross sections reveal a distinct dependence on the polarization of incident $e^-$ and $e^+$ 
beams.\cite{GUTinspired2020c}-\cite{Funatsu2019a}

In this paper we explore the behavior of the model at finite temperature, particularly 
in the context of cosmological evolution of the Universe.
In the SM the electroweak   $SU(2)_L \times U(1)_Y$ symmetry is restored at high temperature.  
In perturbation theory the transition is  weakly first order with $T_c$ around  $160\,$GeV.
We will to show that the behavior of the EW phase transition in GHU is very similar
to that in the SM, though the mechanism of EW symmetry breaking at zero temperature
is quite different.    It will be shown further that a new phase transition, called as the Left-Right (LR) phase 
transition, emerges around $2.5\,$TeV in the GUT inspired GHU.

\section{Effective potential at finite temperature} 

At zero temperature the effective potential $V_\eff (\theta_H, T=0)$ at the one-loop level 
is evaluated from the mass spectra of all fields which depend on $\theta_H$.
It is given by
\begin{align}
V_\eff(\theta_H, T=0) &=
\sum_a  \frac{(-1)^{\eta_a}}{2} \int \frac{d^4 p_E}{(2\pi)^4} \sum_n 
\ln \big\{ p_E^2 + m_n^a (\theta_H)^2 \big\} ~, 
\label{effVgeneral1}
\end{align}
where $\sum_a$ extends over all field multiplets and 
$\eta_a = 0$ or 1 for bosons or fermions, respectively.
When the Kaluza-Klein (KK) spectrum $\{ m_n^a (\theta_H) \}$ is determined by 
the zeros of a function $\rho_a (z; \theta_H)$; namely by
$\rho_a (m_n^a; \theta_H) = 0, ~ (n=1,2,3, \cdots)$, 
then  $V_\eff$ is given  \cite{Falkowski2007} by
\begin{align}
V_\eff (\theta_H, T=0) &= \sum_a 
\frac{(-1)^{\eta_a}}{(4\pi)^2}  \int_0^\infty dy \, y^3 \ln \rho_a (iy; \theta_H) ~.
\label{effVgeneral2}
\end{align}
The $\theta_H$-dependent part of $V_\eff^{1 \, {\rm loop}} (\theta_H)$ is finite,
and independent of the cutoff and regularization method employed.
As explained in the previous section the parameters of the model are determined 
such that $V_\eff (\theta_H, T=0)$ has a global minimum at a desired value of $\theta_H$
and the resultant Higgs boson mass is $m_H = 125.1\,$GeV.

At finite temperature $T \not= 0$, the effective potential becomes
\begin{align}
V_\eff(\theta_H, T) &=
\sum_a   \frac{(-1)^{\eta_a}}{2} \int \frac{d^3 p}{(2\pi)^3} \frac{1}{\beta} \sum_{\ell= -\infty}^\infty \sum_n 
\ln \big\{ \omega_\ell^2 + \vec p {\,}^2 + m_n^a (\theta_H)^2 \big\}  ~, \cr
\noalign{\kern 5pt}
&\hskip 1.cm
\beta = \frac{1}{ T} ~, ~~ \omega_\ell =  \frac{2\pi}{\beta} \Big(\ell +\frac{\eta_a}{2} \Big) ~.
\label{effVfiniteT1}
\end{align}
There appears summation over Matsubara frequencies and 
KK modes.  There are two ways to evaluate it.
One way is to first sum over Matsubara frequencies.  
Employing the identity \cite{Quiros1999}
\begin{align}
&\frac{1}{2} \int \frac{d^3 p}{(2\pi)^3} \frac{1}{\beta} \sum_\ell 
\ln \big\{ \omega_\ell^2 + \vec p {\,}^2 + m^2 \big\} \cr
\noalign{\kern 5pt}
&= \int \frac{d^3 p}{(2\pi)^3} \bigg\{ \frac{1}{2} \omega(p) +
\frac{1}{\beta} \ln \Big(1- (-1)^{\eta}  e^{-\beta \omega(p)} \Big) \bigg\} + \hbox{$m$-independent terms}
\label{effVfiniteT2}
\end{align}
where $\omega(p) = \sqrt{\vec p {\,}^2 + m^2 }$, one finds
\begin{align}
V_\eff(\theta_H, T) &= V_\eff(\theta_H, 0) + \Delta V_\eff(\theta_H, T) ~, \cr
\noalign{\kern 5pt}
\Delta V_\eff(\theta_H, T)  &= \sum_a \sum_n  \frac{(-1)^{\eta_a}}{2 \pi^2 \beta^4} 
\int_0^\infty dx \, x^2 \ln \Big( 1 - (-1)^{\eta_a} e^{-  \sqrt{x^2 + (\beta m_n^a )^2} } \Big)
\label{effVfiniteT3}
\end{align}
where $m_n^a = m_n^a(\theta_H)$.
$\Delta V_\eff(\theta_H, T)$ is finite.  The sum over KK modes converges.
Contributions from modes with $m_n \gg \beta^{-1}=  T$ are negligible.
In the following sections we numerically evaluate $V_\eff(\theta_H, 0)$ by (\ref{effVgeneral2}) 
and $\Delta V_\eff(\theta_H, T)$ by (\ref{effVfiniteT3}).

Alternatively one can evaluate $V_\eff$ by first summing over contributions from the KK modes.
The key observation is that $\sqrt{\omega_\ell^2 + m_n^2}$ in the expression in (\ref{effVfiniteT1})
can be viewed as a mass of the $\ell$th Matsubara mode in $(3+1)$ dimensions.
When the spectrum $\{ m_n (\theta_H) \}$ is determined by 
$\rho (m_n; \theta_H) = 0, ~ (n=1,2,3, \cdots)$, then the spectrum 
$\{ z_n = \sqrt{\omega_\ell^2 + m_n^2} \}$ is determined by 
$\bar \rho (z_n; \theta_H) =\rho (\sqrt{z_n^2 - \omega_\ell^2} \, ; \theta_H) = 0$.
Hence one finds \cite{Hatanaka2012}
\begin{align}
V_\eff(\theta_H, T) &= \sum_a   \sum_\ell 
\frac{(-1)^{\eta_a}}{4\pi^2 \beta}  \int_0^\infty dy \, y^2 \ln \rho_a \big( i\sqrt{y^2 +\omega_\ell^2} \, ; \theta_H \big) ~.
\label{effVfiniteT4}
\end{align}
In  RS space spectrum-determining functions $\rho (z; \theta_H)$ involve Bessel functions
so that the $y$-integral in (\ref{effVfiniteT4}) for each Matsubara mode demands some
time to find the accurate $\theta_H$ dependence of $V_\eff (\theta_H)$.
For this reason we employ the first method using (\ref{effVgeneral2})  and  (\ref{effVfiniteT3})  
to evaluate $V_\eff(\theta_H, T)$ below.
$V_\eff(\theta_H, 0)$ has been already obtained in ref.~\cite{GUTinspired2020b}.

\section{Electroweak phase transition} 

At zero temperature the EW symmetry is dynamically broken by the Hosotani mechanism in GHU.
Dominant contributions to the $\theta_H$-dependent part of $V_\eff(\theta_H, 0)$ come 
from gauge-field multiplets, top-quark multiplet, and dark fermion multiplets at the one-loop level.
In phenomenologically interesting cases  $V_\eff(\theta_H, 0)$ has a global minimum 
around $\theta_H \sim 0.1$ and the KK mass scale $m_\KK$ turns out to be around $10\,$TeV to $15\,$TeV.
In this section we address the question of
when and how the EW symmetry is restored at finite temperature.

In the SM the EW symmetry is spontaneously broken at the tree level, and is restored at finite
temperature.    In perturbation theory the transition occurs 
at $T_c^\EW \sim 160\,$GeV, and is  weakly first order.\cite{Quiros1999, Senaha2020}
In the lattice simulation the transition is observed to be smoother.\cite{Onofrio2016}
Although the EW symmetry breaking mechanism at $T=0$ in GHU is quite different from that in the SM, 
the behavior of the EW symmetry restoration at the weak scale  is expected to besimilar.
In GHU $T_c^\EW \ll m_\KK$ so that only SM particles are expected to give relevant contributions to 
$\Delta V_\eff(\theta_H, T)$ in (\ref{effVfiniteT3}) at $T \lesssim T_c^\EW$.

\begin{figure}[tbh]
\centering
\includegraphics[width=100mm]{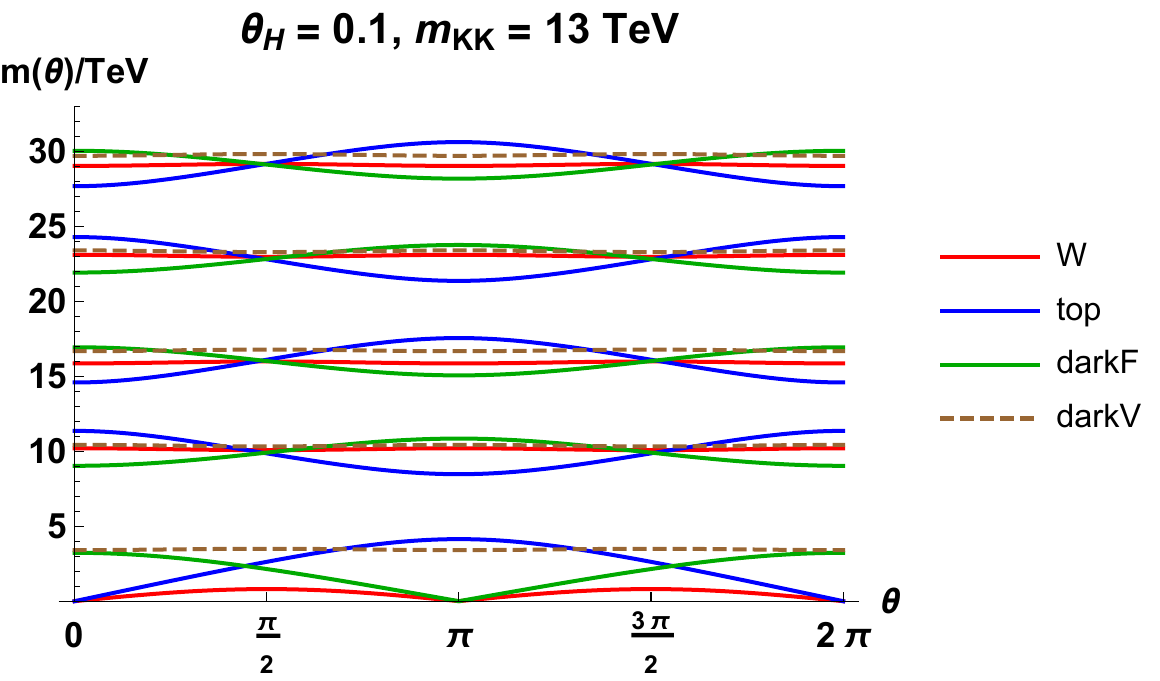}
\caption{Mass spectra for $W$ boson ($W^{(n)}, \hat W^{(n)}$), top ($t^{(n)}$ and $t^{\prime (n)}$), darkF and 
darkV towers when  $V_\eff (\theta_H, 0)$  has a global minimum at $\theta_H^{\rm min}=0.1$, and other 
parameters are given by $m_\KK = 13\,$TeV and $(n_F, n_V, c_V)=(2,4, 0.2)$.
For the $W$ series levels from the bottom are $W^{(0)}, \hat W^{(1)}, W^{(1)}, \hat W^{(2)}, W^{(2)}, \cdots$,
while for the top series they are $t^{(0)}, t^{\prime (1)}, t^{(1)}, t^{\prime (2)}, t^{(2)}, \cdots$.
}   
\label{fig:spectrum1}
\end{figure}

Recall that only KK towers with $\theta_H$-dependent $m_n(\theta_H)$ are relevant to
$\Delta V_\eff(\theta_H, T)$ in (\ref{effVfiniteT3}).  
They are $W$ tower, $Z$ tower, $A_z$ (Higgs) tower, top quark tower, bottom quark tower,
dark fermion $\Psi_F$ (darkF) tower, and dark fermion $\Psi_V$ (darkV) tower.  
The spectrum-determining $\rho (z; \theta_H)$ functions are tabulated in Appendix A of
ref.\ \cite{GUTinspired2020b}.
Other  quark  and lepton multiplets have $\theta_H$-dependent spectra $m_n (\theta_H)$,
but the magnitude of their $\theta_H$-dependence is
small and almost irrelevant to $\Delta V_\eff(\theta_H, T)$.
The spectra $\{ m_n^a (\theta) \}$ for $W$, top, darkF and darkV towers are displayed 
in Fig.\ \ref{fig:spectrum1}.

The mass spectrum of the top quark has the largest $\theta_H$ dependence.
The spectrum of the  darkF has the second largest $\theta_H$ dependence.  
As opposed to the top-quark case, the  darkF is massive at $\theta_H = 0$ while it
becomes massless at $\theta_H = \pi$.
The spectrum of the darkV tower has much weaker $\theta_H$ dependence.  Although it is important
at zero temperature, it gives little effect for the behavior of $V_\eff (\theta_H, T)$ at  finite temperature.

\begin{figure}[tbh]
\centering
\includegraphics[width=110mm]{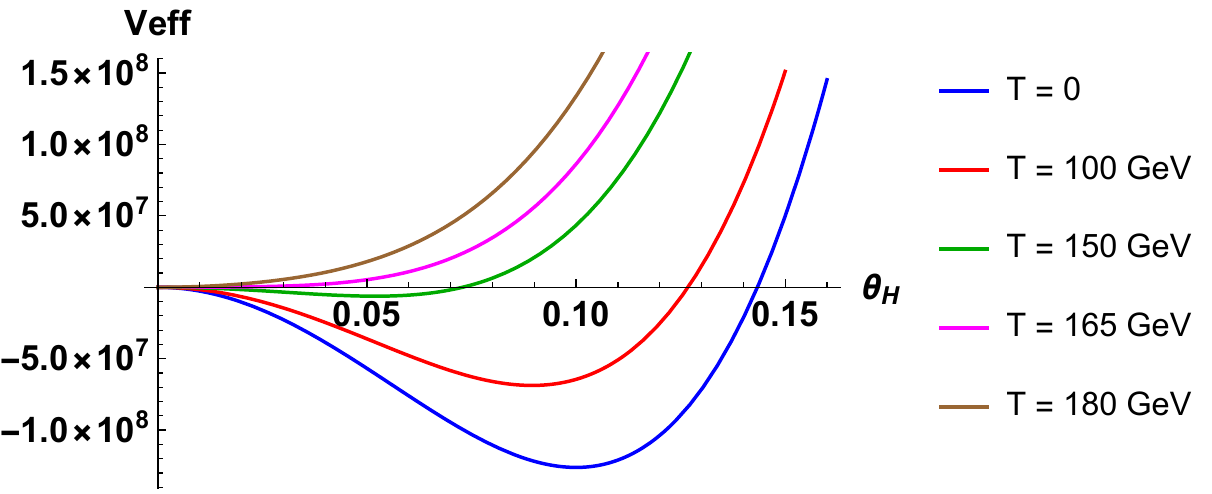}
\caption{Behavior of $V_\eff (\theta_H, T)$ in units of GeV$^4$ for $T=0 \sim 180\,$GeV.  
At $T=0$ $V_\eff$ has a minimum at $\theta_H^{\rm min}=0.1$. 
}   
\label{fig:EWtransition1}
\end{figure}

\begin{figure}[tbh]
\centering
\includegraphics[width=80mm]{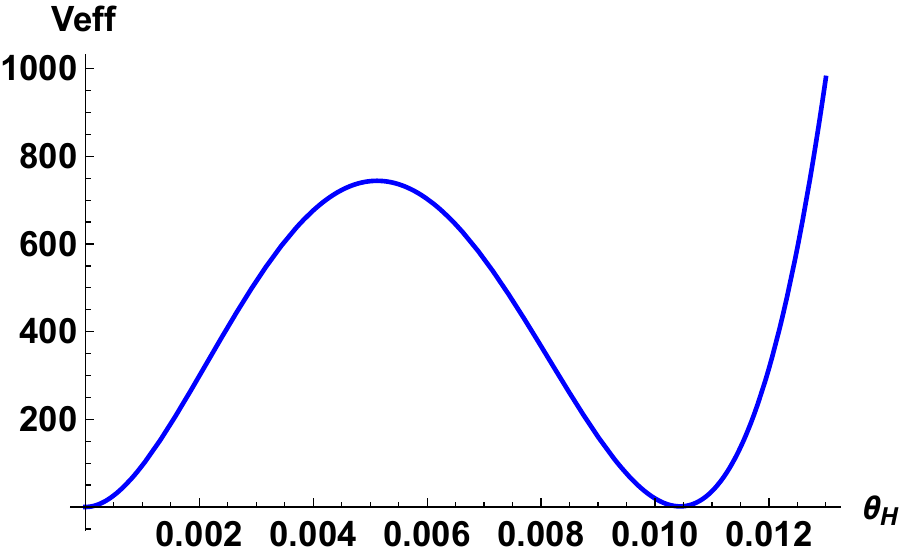}
\caption{$V_\eff (\theta_H, T_c^\EW)$ in units of GeV$^4$ is plotted 
at $T= T_c^\EW = 163.2\,$GeV  for $(\theta_H^{\rm min}, n_F, n_V, c_V)= (0.1, 2, 4, 0.2)$.  
Degenerate minima are located at $\theta_H=0$ and 
$\theta_H^c= 0.0104$.
$v_c= \theta_H^c f_H = 25.54\,$GeV so that $v_c/T_c^\EW  = 0.156$.
}   
\label{fig:EWtransition2}
\end{figure}

We insert those mass spectra $\{ m_n^a (\theta) \}$ into  (\ref{effVfiniteT3}) to find $V_\eff (\theta_H, T)$.
Its behavior for $0\,{\rm GeV} \le T \le 180\,$GeV and $0 \le \theta_H \le 0.15$ is depicted in Fig.~\ref{fig:EWtransition1}.
Here and below $V_\eff (\theta_H, T) - V_\eff (0, T)$ has been plotted in figures.
For $T < 300\,$GeV, only contributions from the SM fields, namely $W$, $Z$, Higgs, and top-quark fields,
are relevant to $\Delta V_\eff (\theta_H, T)$.  The EW symmetry is restored around 163$\,$GeV.
Near the critical temperature one needs careful evaluation. 
As in the SM  a small bump develops for small $\theta_H$ as a result of $T |\phi|^3$-type
contributions from bosons.  When $V_\eff (\theta_H, 0)$ has a global minimum at
$\theta_H= \theta_H^{\rm min}=0.1$ and other parameters are given by $m_\KK = 13\,$TeV, $n_F=2, n_V=4$ and 
$c_V= 0.2$, the critical temperature is found to be $T_c^\EW = 163.2\,$GeV.
$V_\eff (\theta_H, T_c^\EW)$ for $0 \le \theta_H < 0.013$ is evaluated numerically and is 
depicted in Fig.~\ref{fig:EWtransition2}.  The degenerate minimum is located
at $\theta_H^c = 0.0104$ and $v_c = \theta_H^c f_H = 25.54\,$GeV. 
The ratio $v_c/T_c^\EW$  is 0.156.  The transition is  weakly first order.
It has been known in the SM  that  $T_c^{\EW,\SM} \simeq 163.4\,$GeV, 
$v_c^\SM \simeq 24.3\,$GeV
and $v_c^\SM/T_c^{\EW,\SM} \simeq 0.15$ in the one-loop approximation.\cite{Senaha2020}
Although the EW symmetry breaking mechanism in GHU is quite different from that in the SM,
the behavior at finite temperature ($T < 300\,$GeV) in GHU is almost the same as in the SM.

We remark that this behavior does not depend on  detailed values of various parameters in the model. 
As mentioned above, there are a few parameters in the dark fermion sector which can be taken differently.  
One finds that $T_c^\EW = 163.2\,$GeV for $(\theta_H^{\rm min}, n_F, n_V, c_V)= (0.1, 2, 4, 0.2)$, and  
$T_c^\EW = 163.3\,$GeV for  $ (0.1, 5, 2, 0.2)$.   
The $\theta_H$ universality remains valid for the quantity $T_c^\EW$.
$T_c^\EW$  depends on $\theta_H^{\rm min}$ very little, too.  One finds, for instance,  that
$T_c^\EW = 163.3\,$GeV for $(\theta_H^{\rm min}, n_F, n_V, c_V)= (0.11, 2, 4, 0.2)$.
All evaluations of the critical temperature in this section have been carried out 
at the one-loop level.  Higher-order corrections may affect the values just as in the SM.

\section{Left-right phase transition} 

\begin{figure}[tbh]
\centering
\includegraphics[width=120mm]{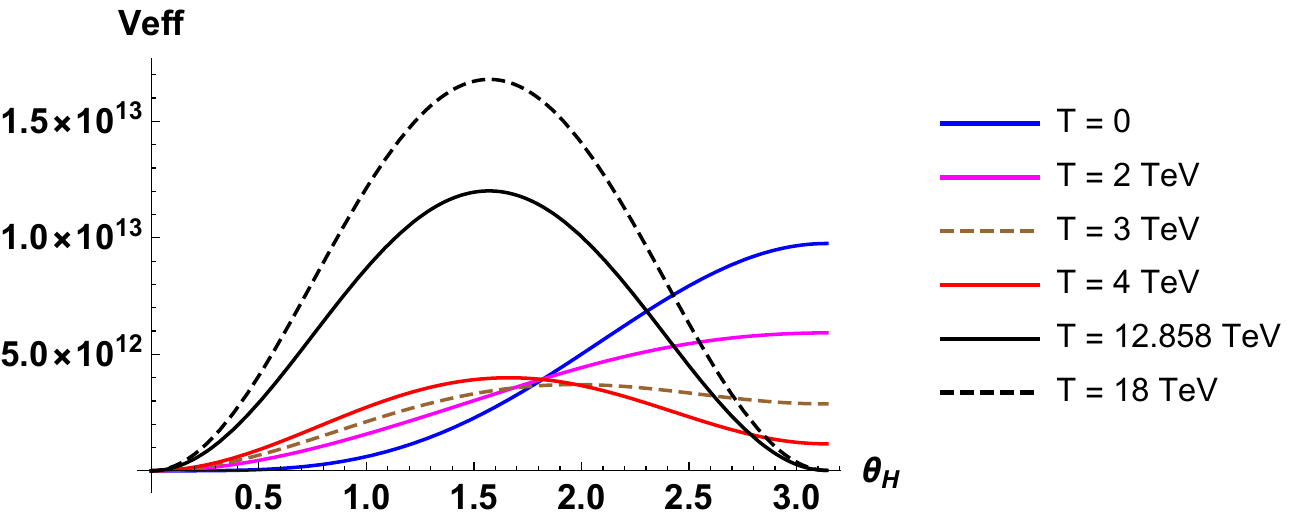}
\caption{Behavior of $V_\eff (\theta_H, T)$ in units of GeV$^4$ for $T=0 \hbox{ -- } 18\,$TeV.  
$\theta_H^\min = 0.1$, $m_\KK = 13\,$TeV, and $(n_F, n_V, c_V) = (2,4, 0.2)$.}   
\label{fig:LRtransition1}
\end{figure}

As the temperature is raised further, a new feature emerges in the global behavior of
$V_\eff (\theta_H, T)$.  $\theta_H = \pi$ becomes a local minimum
of $V_\eff (\theta_H, T)$ at $T= T_{c2}^\LR \sim 2.3\,$TeV, and becomes a global minimum 
at $T=T_{c1}^{\LR} \sim m_\KK$.
Its behavior is plotted in Fig.~\ref{fig:LRtransition1}.

In  expression  (\ref{effVfiniteT3}) of $V_\eff (\theta_H, T)$ the contributions from $W$, $Z$, Higgs, and darkV
towers are periodic in $\theta_H$ with a period $\pi$, giving the same amount of contributions at $\theta_H=0$ and $\pi$.
On the other hand contributions from top-quark and darkF towers have periodicity  with a period $2\pi$, giving rise to
a difference between $\theta_H=0$ and $\pi$.  Furthermore, the top quark is massless and the darkF is massive at $\theta_H=0$,
whereas the top quark is massive and the darkF is massless at $\theta_H=\pi$.  
In effect, the role of top quark and darkF is interchanged.

As $T$ is increased further above $m_\KK$, it is expected from  (\ref{effVfiniteT4}) that contributions 
from boson fields dominate over those from fermion fields.  For fermions the  Matsubara frequency $|\omega_\ell|$
is equal to or larger than $\pi T$, whereas for bosons there exist zero frequency modes $\omega_0 = 0$.  
Fermion contributions are suppressed compared to boson contributions, which in turn implies, in the current case, 
that $\theta_H=0$ and $\theta_H=\pi$ phases become
almost degenerate at sufficiently high temperature.

This leads to an important consequence in the history of the evolution of the early Universe.
As the Universe expands and the temperature drops to $T \sim m_\KK$, the $\theta_H=0$ and $\theta_H=\pi$
states are almost degenerate so that the Universe would settle in the domain structure.
As is seen below, the $\theta_H=\pi$ state remains stable until $T$ drops further to $T_\decay^{\LR} \sim 2.6\,$TeV
at which time tunneling from $\theta_H=\pi$ to $\theta_H=0$ rapidly takes place.

We shall see  in section 6 that the role of $SU(2)_L$ and $SU(2)_R$ is  
interchanged in the $\theta_H=0$ and $\theta_H=\pi$ states.  For this reason the 
$\theta_H= 0 \leftrightarrow \pi$ transition is called  the Left-Right (LR) transition.

\subsection{Critical temperatures $T_{c1}^{\LR}$ and $T_{c2}^{\LR}$}

There is a critical temperature  $T_{c1}^{\LR}$ at which 
$V_\eff (0, T_{c1}^{\LR}) = V_\eff (\pi, T_{c1}^{\LR})$.
For $T > T_{c1}^\LR$, $V_\eff (0, T)  >  V_\eff (\pi, T)$.
It should be remembered that $V_\eff (0, T)  -  V_\eff (\pi, T) \ll V_\eff (\onehalf \pi, T)  -  V_\eff (\pi, T)$
for $T > T_{c1}^\LR$.  
There is another critical temperature $T_{c2}^\LR $.
While the $\theta_H = \pi$ state remains as a local minimum for  $T_{c2}^\LR < T < T_{c1}^\LR$, 
 the $\theta_H = \pi$ state becomes a maximum
of $V_\eff (\theta_H, T)$ for $T < T_{c2}^\LR$, hence becoming absolutely unstable.

To find the values of $T_{c1}^{\LR}$ and $T_{c2}^{\LR}$ one need to sum over the contributions from a large number of 
KK modes in (\ref{effVfiniteT3}).  Since only top quark and darkF towers are relevant for this quantity,
one can write, for $V_\eff^\LR (T) \equiv V_\eff(\pi, T) - V_\eff(0, T)$, 
\begin{align}
&V_\eff^\LR (T) = V_\eff^\LR (0)+ \delta V_\eff^\LR (T) ~, \cr
\noalign{\kern 5pt}
& \delta V_\eff^\LR (T)  \simeq 
{\sum_a}'  \sum_{n=1}^{n_\max}  \frac{-1}{2 \pi^2 \beta^4} 
\int_0^\infty dx \, x^2 \ln \frac{1+ e^{ -\sqrt{x^2 + [\beta m_n^a (\pi) ]^2} }}{1+ e^{ -\sqrt{x^2 + [\beta m_n^a (0)]^2} }} ~,
\label{LRV1}
\end{align}
where the sum $\sum_a'$ extends over top quark and darkF towers.

\begin{figure}[tbh]
\centering
\includegraphics[width=90mm]{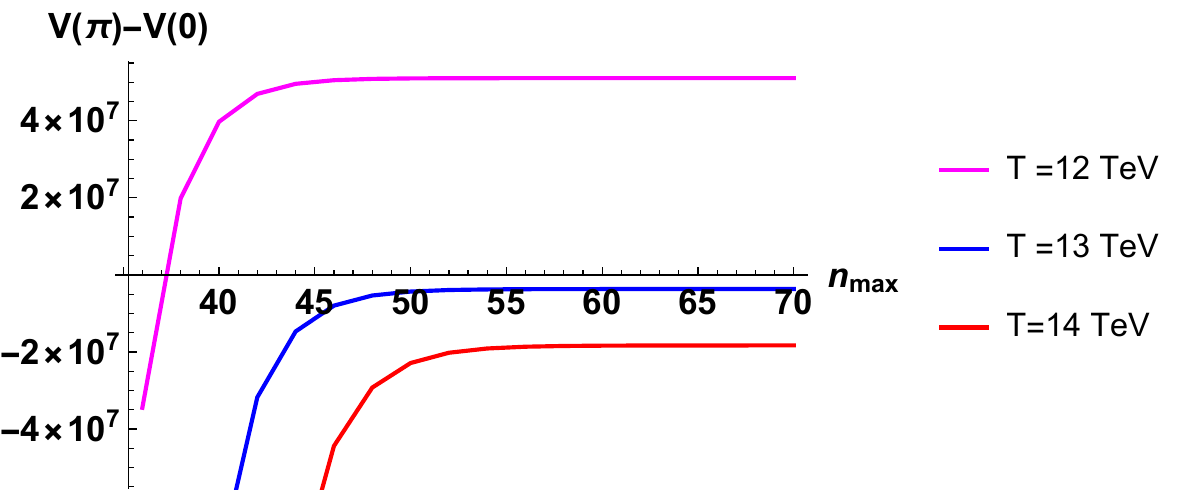}
\caption{$V_\eff^\LR (T) $ in (\ref{LRV1}) in  units of GeV$^4$  is plotted with varying 
an even integer $n_\max$.  
$m_\KK = 13\,$TeV, $\theta_H^\min = 0.1$, and $(n_F, n_V, c_V) = (2, 4, 0.2)$.}   
\label{fig:LRtransition2}
\end{figure}

For $T > m_\KK$ a large number of KK modes contribute.  
As seen from the spectrum depicted in Fig.~\ref{fig:spectrum1}, the behavior of the $\theta_H$-dependence of 
$m_n^a (\theta_H)$ alternates as $n$.  There results partial cancellation between the $n$th mode 
and $(n+1)$th mode in the formula (\ref{LRV1}).  
$V_\eff^\LR (T)$ in (\ref{LRV1})  is plotted in Fig.~\ref{fig:LRtransition2}   
with an even integer $n_\max$ varied.  
One can see that $n_\max \ge 50$ is necessary near the critical temperature to reach an asymptotic value.

The critical temperature $T_{c1}^{\LR}$ turns out to be very close to $m_\KK$,
and has little dependence on $\theta_H^{\min}$.
The other critical temperature $T_{c2}^{\LR}$ turns out to be around 2.3$\,$TeV.
It is tabulated in Table \ref{Tab:LRtransition1}  with various choices of the parameters.

\begin{table}[tbh]
\renewcommand{\arraystretch}{1.1}
\begin{center}
\vskip 10pt
\begin{tabular}{|c|c|c|c|c||c|c|c|c|}
\hline
$\theta_H^{\min}$ & $m_\KK $ & $n_F$&$n_V$ &$c_V$ &$c_F$ &$m_V/k$ &$T_{c1}^{\LR}$ &$T_{c2}^{\LR}$\\
\hline
0.1&13 TeV &2 & 4 & 0.2 & 0.358042 & 0.086414 &12.86 TeV &2.348 TeV \\
\cline{3-9}
& &5 & 2 & 0.2 & 0.456079 & 0.071245 &12.85 TeV &2.238 TeV\\
\cline{2-9}
 &11 TeV &2 & 4 & 0.2 & 0.236826& 0.106592 &10.88 TeV &2.277 TeV\\
 \hline
0.11&13 TeV &2 & 4 & 0.2 & 0.392398& 0.075104&12.86 TeV &2.215 TeV\\
\hline
\end{tabular}
\caption{\small
$T_{c1}^{\LR}$ and $T_{c2}^{\LR}$. $(\theta_H^{\min}, m_\KK, n_F, n_V, c_V)$ are input
parameters.
}
\label{Tab:LRtransition1}
\end{center}
\end{table}

\subsection{Bounce solutions and $T_\decay^{\LR}$}

Although $V_\eff(0, T) < V_\eff(\pi, T)$ for $T < T_{c1}^{\LR}$, the transition from the $\theta_H=\pi$
phase to the $\theta_H=0$ phase does not proceed immediately.  The temperature must drop further
before a rapid transition takes place. We need to evaluate bounce solutions at finite temperature
to estimate the tunneling rate.

As shown in Fig.~\ref{fig:LRtransition1} in the case of $\theta_H^\min = 0.1$, $m_\KK=13\,$TeV and
$(n_F, n_V, c_V) = (2,4, 0.2)$, the $\theta_H = \pi$ state is at a local minimum of $V_\eff (\theta_H, T)$
at $T=4\,$TeV.  The tunneling rate per unit time per unit volume is given 
in the form $A(T) e^{-S_3 /T}$ where $S_3$ is the three-dimensional action of a bounce solution;\cite{Quiros1999}
\begin{align}
S_3 &= \int d^3 x \, \bigg\{ \frac{1}{2} (\nabla \phi )^2 + V_\eff  \Big( \frac{\phi}{f_H} , T \Big) 
- V_\eff ( \pi , T) \bigg\} ~,  \cr
\noalign{\kern 5pt}
&\frac{d^2 \phi}{dr^2} + \frac{2}{r} \frac{d\phi}{dr} 
- \frac{1}{f_H} V_\eff^{(1)} \Big(\frac{\phi}{f_H} , T \Big) = 0~, \cr
\noalign{\kern 5pt}
&\lim_{r \go \infty} \phi(r) = \pi f_H ~, \cr
\noalign{\kern 5pt}
&\frac{d\phi}{dr} \Big|_{r=0} = 0 ~.
\label{bounce1}
\end{align}
In terms of dimensionless quantities $\theta = \phi/f_H$, $t = f_H r$, and 
$U(\theta, T) = - f_H^{-4} \, V_\eff(\theta, T)$,  we have
\begin{align}
&S_3 =4\pi f_H \int_0^\infty dt \,  t^2 \bigg\{ \frac{1}{2} \Big( \frac{d\theta}{dt} \Big)^2  
- U(\theta,T) + U(\pi,T) \bigg\} ~,  \cr
\noalign{\kern 5pt}
&\frac{d^2 \theta}{dt^2} + \frac{2}{t} \frac{d\theta}{dt} +  \frac{dU}{d\theta} = 0~, 
\label{bounce2}
\end{align}
with conditions $d\theta/dt|_{t=0}=0$ and $\theta |_{t=\infty} = \pi$.
The problem is reduced to determining the motion of a particle in a potential $U$.

\begin{figure}[tbh]
\centering
\includegraphics[width=110mm]{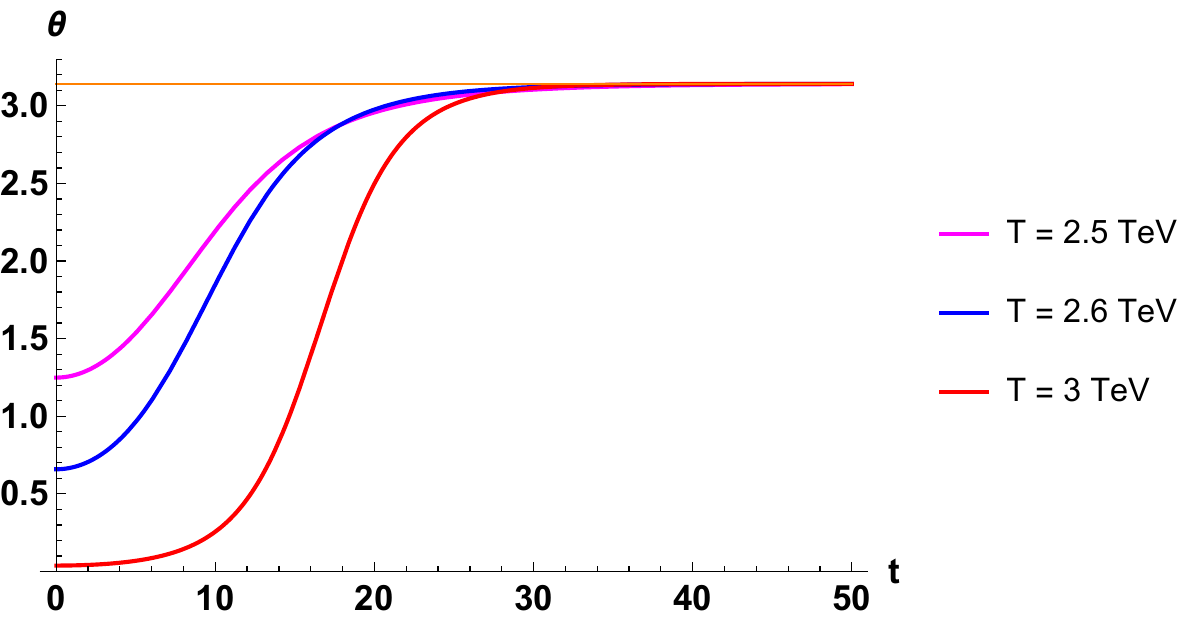}
\caption{Bounce solutions $\theta (t)$, where $t= f_H r$,  for $T=2.5\,$TeV, $2.6\,$TeV and $3\,$TeV
in the case  of $\theta_H^\min = 0.1$, $m_\KK = 13\,$TeV   and $(n_F, n_V, c_V) = (2, 4, 0.2)$.}
\label{fig:bounce1}
\end{figure}

Bounce solutions can be easily found.  Solutions for $T=2.5\,$TeV,  $T=2.6\,$TeV and $3\,$TeV are
displayed in Fig.~\ref{fig:bounce1}.
For higher temperatures, say, $T=4\,$TeV, $\theta (0)$ must be very close to 0, in which case
the thin-wall approximation becomes legitimate.

For the tunneling rate the most relevant  quantity is $S_3/T$.   
The result is summarized in Table \ref{Tab:LRbounce}.
Bubble nucleation rate becomes sufficiently large so that
the LR transition from the $\theta_H = \pi$ state to the $\theta_H = 0$ state rapidly proceeds
at $T = T_\decay^{\LR}$ 
when $S_3/T$ becomes $O(130 \sim 140)$.\cite{McLerran1991, Anderson1992, Dine1992, Moreno1998}
It is seen that in the case of $m_\KK = 13\,$TeV, $T_\decay^{\LR} \sim 2.6\,$TeV for $\theta_H^\min = 0.1$
and $T_\decay^{\LR} \sim 2.45\,$TeV for $\theta_H^\min = 0.11$, respectively.

\begin{table}[tbh]
\renewcommand{\arraystretch}{1.1}
\begin{center}
\begin{tabular}{|c|c|c|c|c|}
\hline
$\theta_H^{\min}$ & $m_\KK $ & $T$ &$\theta(0)$ &$S_3/T$ \\
\hline
0.1&13 TeV&3 TeV & $0.0379$ & 616  \\ 
\cline{3-5}
& &2.65 TeV & $0.4761$ & 190  \\ 
\cline{3-5}
& &2.6 TeV & $0.6585$ & 150  \\ 
\cline{3-5}
& &2.55 TeV & $0.9094$ & 112  \\ 
\cline{3-5}
& &2.5 TeV & $1.2490$ & 77  \\ 
\cline{2-5}
&11 TeV &3 TeV & $0.0791$ & 1107  \\ 
\cline{3-5}
& &2.5 TeV & $0.7000$ & 160   \\ 
\cline{3-5}
& &2.45 TeV & $1.0050$ & 113  \\ 
\cline{3-5}
& &2.4 TeV & $1.4246$ & 71  \\ 
\hline
0.11&13 TeV&3 TeV & $0.0864$ & 840  \\ 
\cline{3-5}
& &2.55 TeV & $0.3490$ & 212  \\ 
\cline{3-5}
& &2.5 TeV & $0.4875$ & 171  \\ 
\cline{3-5}
& &2.45 TeV & $0.6802$ & 133   \\ 
\cline{3-5}
& &2.4 TeV & $0.9488$ & 98 \\ 
\hline
\end{tabular}
\caption{\small
$S_3/T$ for bounce solutions.  $(n_F, n_V, c_V) = (2, 4, 0.2)$.
Initial values $\theta (0)$ are tuned to yield bounce solutions.
}
\label{Tab:LRbounce}
\end{center}
\end{table}

The $\theta_H=\pi$ state corresponds to $v_c = \pi f_H = 7.71\,$TeV for $\theta_H^\min = 0.1$ 
and $m_\KK = 13\,$TeV,  which gives $v_c/T_\decay^{\LR} \sim 2.97$. 
One might wonder whether or not gravitational waves (GWs) generated in the LR phase transition 
can be detected in future GW observations.
There are two relevant quantities denoted as $\alpha$ and $\beta$ in the literature 
for describing dynamics of a first-order phase transition in association with generations of 
GWs.\cite{Kamionkowski1994, Nicolis2004, Kakizaki2015} 
$\alpha$ is  the ratio between the false vacuum energy (latent heat) density and the  thermal energy density
at $T_\decay^\LR$, which gives  a measure of the transition strength.
$\beta$ is the rate of time variation of the nucleation rate at the transition.
The number $g_*$ of relativistic degrees of freedom at $T_\decay^\LR$ is $96.25 + 42 n_F$ 
(180.25 for $n_F = 2$).  We have found that $\alpha \sim 0.004$ and $\beta / H_* \sim 2100$ 
in the case $\theta_H^\min=0.1, m_\KK=13\,$TeV and $(n_F, n_V, c_V)=(2,4, 0.2)$, 
where $H_*$ is the Hubble parameter at the transition.
The amount of energy released in the LR transition is small, giving a tiny value of $\alpha$.
A GW signal from the LR transition is far below the reach of the sensitivity of, say, 
LISA.

Before closing this section we summarize the cosmological history of the Universe in Table \ref{Tab:history}
after the temperature drops around $T=m_\KK$.
As remarked in the introduction, the scenario is valid as long as the RS warped space is stable around $T_\decay^\LR$.  
If the Universe is in a  decompactified phase discussed in refs.\ \cite{Creminelli2002, Agashe2020}
at $T  <  m_\KK$,  the scenario may need to be modified accordingly.

\begin{table}[tbh]
\renewcommand{\arraystretch}{1.3}
\begin{center}
\begin{tabular}{cl}
Temperature&\qquad Phase of the Universe  \\
\hline \hline
$m_\KK \sim 13\,$TeV & :~Domain structure of $\theta_H=0$ and $\theta_H=\pi$ phases is formed. \\
\hline
$\downarrow$&\quad  domains consisting of 
$\begin{cases}\theta_H=0 ~:~ SU(2)_L \times U(1)_Y ~ {\rm phase} \cr
\theta_H=\pi ~:~ SU(2)_R \times U(1)_{Y'} ~ {\rm phase} \end{cases}$\\
\hline
$T_\decay^\LR \sim 2.5\,$TeV& :~LR first-order transition from $\theta_H=\pi$ phase to $\theta_H=0$ phase\\
\hline
$\downarrow$&\qquad $\theta_H=0 ~:~ SU(2)_L \times U(1)_Y$ phase\\
\hline
$T_c^\EW \sim 163\,$GeV &:~EW weakly first-order transition to $\theta_H \not= 0$\\
\hline
$\downarrow$&\qquad $\theta_H \not= 0 ~:~ SU(2)_L \times U(1)_Y$ is broken to $U(1)_\EM$\\
\hline
$T\sim 0$ &:~The present universe\\
\hline
\end{tabular}
\caption{
Schematic view of the history of the Universe is shown.}
\label{Tab:history}
\end{center}
\end{table}

\section{Gauge symmetry and couplings at  $\theta_H=0$ and $\pi$} 

In the previous section we have seen that the Universe forms domain structure above $T= T_\decay^{\LR}$
in which the $\theta_H=0$   and $\theta_H=\pi$  states coexist.
The $\theta_H=0$  state is the $SU(2)_L \times U(1)_Y$ symmetric phase.
It is important to understand what the $\theta_H=\pi$  state is.

In this section it is shown that  the $\theta_H=\pi$  state corresponds to a state with
$SU(2)_R \times U(1)_{Y'}$ symmetry, which becomes manifest in the twisted gauge.
In $SO(5) \times U(1)_X$ GHU the $SU(2)_L \times U(1)_Y$ and 
$SU(2)_R \times U(1)_{Y'}$ phases are connected smoothly by $\theta_H$.
Mass spectrum and gauge couplings of quarks and leptons continuously change as $\theta_H$ varies 
from 0 to $\pi$.   To find those gauge couplings,  wave functions of gauge fields and fermion fields in the fifth
dimension must be first determined.  Details of  wave functions are given  in Appendixes.

\subsection{Twisted gauge}

When $V_\eff (\theta_H, T=0)$ is minimized at $\theta_H \not= 0$, the EW symmetry is spontaneously 
broken to $U(1)_\EM$ in general.   It has been known that gauge couplings and other physical quantities
in the vacuum with $\theta_H \not= 0$  can be most conveniently evaluated in the twisted 
gauge. This remains valid at $T\not= 0$.

In GHU one can make a large gauge transformation such that the AB phase $\theta_H$ defined 
in (\ref{ABphase2}) becomes zero in a new gauge.\cite{Falkowski2007, HS2007} 
To be more explicit, consider an $SO(5)$  gauge 
transformation
\begin{align}
&\tilde A_M = \tilde \Omega A_M  \tilde \Omega^{-1} + \frac{i}{g_A} \tilde \Omega \dd_M \tilde \Omega^{-1} ~, \cr
\noalign{\kern 5pt}
&\tilde \Omega = e^{i \theta (z) T^{45}} ~, ~~ \theta (z) = \theta_H \, \frac{z_L^2 - z^2}{z_L^2 - 1} ~.
\label{twist1}
\end{align}
As $\la \tilde A_z \ra  = \la   A_z \ra + g_A^{-1} \theta' (z) T^{45}$, the AB phase in the new gauge 
becomes $\tilde \theta_H = 0$.  This gauge is called   the twisted gauge.
Quantities in the twisted gauge are denoted with a tilde  in this section and Appendixes.
Although the AB phase vanishes, orbifold boundary conditions are changed to
\begin{align}
&\tilde P_0^{\rm vec} = 
\begin{pmatrix} 1 &&&& \cr &1 &&&\cr  && 1 && \cr 
&&& \cos 2 \theta_H & - \sin 2 \theta_H \cr &&& -\sin 2 \theta_H & - \cos 2 \theta_H  \end{pmatrix},~~
\tilde P_1^{\rm vec} =  \begin{pmatrix} 1 &&&& \cr &1 &&&\cr  && 1 && \cr  &&&1 &\cr &&&&-1 \end{pmatrix},\cr
\noalign{\kern 5pt}
&\tilde P_0^{\rm sp} = \sigma^0 \otimes (\cos \theta_H \sigma^3 + \sin \theta_H \sigma^2) , ~~
\tilde P_1^{\rm sp} = \sigma^0 \otimes \sigma^3 .
\label{twist2}
\end{align}
Here $\tilde P_j^{\rm vec}$ and $\tilde P_j^{\rm sp}$ represent orbifold boundary condition matrices
in the vectorial and spinorial representations, respectively.
In the twisted gauge, boundary conditions at $z=1$ becomes nontrivial and $\theta_H$ dependent, 
but equations and wave functions of various fields in the bulk ($1 < z \le z_L$) become simple 
as the background field $\tilde \theta_H$ vanishes.  Physics does not depend on the gauge.

$SO(5)$ gauge fields are decomposed as
\begin{align}
&A_M = \frac{1}{\sqrt{2}} \sum_{1 \le j < k \le 5} A_M^{(jk)} T^{jk} ~, \cr
\noalign{\kern 5pt}
&[ T^{jk} ,  T^{\ell m}] = i (\delta^{j\ell} T^{k m} - \delta^{j m} T^{k \ell}
- \delta^{k \ell} T^{j m} + \delta^{k m} T^{j \ell}) ~.
\label{SO5gauge1}
\end{align}
Four-dimensional components in the twisted gauge are given by
\begin{align}
&\tilde A_\mu = \frac{1}{\sqrt{2}} \sum_{1 \le j < k \le 5} \tilde A_\mu^{(jk)} T^{jk} 
= \tilde \Omega A_\mu  \tilde \Omega^{-1} ~.
\label{SO5gauge2}
\end{align}
Generators of $SU(2)_L$ and $SU(2)_R$ are
\begin{align}
&\begin{pmatrix} T^{a_L} \cr T^{a_R} \end{pmatrix} 
= \frac{1}{2} \Big( \frac{1}{2} \ep^{abc} T^{bc}  \pm T^{a4} \Big) , ~~ a,b,c = 1 \sim 3 ~.
\label{SO5generator1}
\end{align}
To investigate the relation between the original and twisted gauges, 
let us define $\tilde T^{jk} (z) = \tilde \Omega T^{jk} \tilde \Omega^{-1}$.  It follows that
\begin{align}
\left[ \begin{matrix} \tilde T^{a 4} (z)  \cr \tilde T^{a 5} (z) \end{matrix} \right]&= 
\left[ \begin{matrix} \cos \theta (z) & - \sin \theta (z)  \cr \sin \theta (z)  & \cos \theta (z) \end{matrix} \right]
\left[ \begin{matrix}  T^{a 4}   \cr  T^{a 5} \end{matrix} \right]
\quad (a=1,2,3)
\label{SO5generator2}
\end{align}
and other components remain unchanged.
Recall that $\theta (1) = \theta_H$ and $\theta (z_L) = 0$.
In particular for $\theta_H = \pi$,  $\tilde T^{a4} (1) = - T^{a4}$ and $\tilde T^{a4} (z_L) = + T^{a4}$.
In the basis of $\{ \tilde T^{jk} (z) \}$ the role of $T^{a_L}$ and $T^{a_R}$ is interchanged 
as $z$ varies from $z=1$ to $z_L$.
Indeed this property becomes crucial in discussing gauge symmetry in the $\theta_H=\pi$ state.

\subsection{Gauge symmetry and couplings}
\def\myspace{\noalign{\kern 3pt}}

Wave functions of KK towers of gauge fields in the twisted gauge are given in Appendix B.
With $A_\mu^{a_{L/R}} = 2^{-1/2} (\onehalf \ep^{abc} A_\mu^{(bc)} \pm  A_\mu^{(a4)})$
and $A_\mu^{\hat a} = A_\mu^{a 4}$, 
a set $(A_\mu^{b_L}, A_\mu^{b_R}, A_\mu^{\hat b})$ ($b=1,2$) forms  charged gauge-field towers, 
which are  decomposed into $W$, $\hat W$, and $W_R$  towers.
Mass eigenvalues are given by $\{ k \lambda_n \}$ in each KK tower.  They are determined by
\begin{align}
&\hbox{$W$ tower:} \quad 2 S C'(1; \lambda_{W^{(n)}}) + \lambda_{W^{(n)}} \sin^2 \theta_H = 0 ~, \cr
&\hbox{$\hat W$ tower:} \quad 2 S C'(1; \lambda_{\hat W^{(n)}}) + \lambda_{\hat W^{(n)}} \sin^2 \theta_H = 0 ~, \cr
&\hbox{$W_R$ tower:} \quad  C(1; \lambda_{W_R^{(n)}})  = 0 ~.
\label{Wtower2}
\end{align}
where functions $C(z; \lambda), S(z; \lambda)$, etc.\  are defined in (\ref{functionA1}).
The spectra of the $W$ and $\hat W$  towers for $\sin \theta_H=0$ reduce to those determined by
$C'(1; \lambda_{W^{(n)}})  = 0$ and $S(1; \lambda_{\hat W^{(n)}})  = 0$, respectively.
The $W$ tower has a zero mode $\lambda_{W^{(0)}}$ for $\sin \theta_H=0$, whereas the $\hat W$  tower
does not.  As depicted in Fig.\ \ref{fig:spectrum1} the spectra of the $W$ and $\hat W$  towers alternate.
The lowest level [$W^{(0)}$]  corresponds to $W$ boson.

For $\theta_H=0$ and $\pi$, the $W$ boson becomes massless, but the property of the $W$ boson at $\theta_H=\pi$
is quite different from that at $\theta_H=0$.  The $W$ boson at $\theta_H=0$ is a purely $SU(2)_L$ gauge boson.
As is manifest from the expression in (\ref{Wtower1}), $W_\mu^{(n)} (x) $ fields are purely $SU(2)_R$ 
at $\theta_H=\pi$ in the twisted gauge.   

In the sector of neutral gauge bosons, $(A_\mu^{3_L}, A_\mu^{3_R}, A_\mu^{\hat 3}, B_\mu)$, 
there are $Z$, $\hat Z$, $Z_R$ and $\gamma$ towers.  (A neutral tower from $A_\mu^{\hat 4}$ 
does not couple with quarks and leptons.)
Mass eigenvalues  in each KK tower are determined by
\begin{align}
\hbox{$Z$ tower:} \quad &2 S C'(1; \lambda_{Z^{(n)}}) + 
(1 + s_\phi^2) \lambda_{Z^{(n)}} \sin^2 \theta_H = 0 ~, \cr
\hbox{$\hat Z$ tower:} \quad &2 S C'(1; \lambda_{\hat Z^{(n)}}) 
+ (1 + s_\phi^2) \lambda_{\hat Z^{(n)}} \sin^2 \theta_H = 0 ~, \cr
\hbox{$Z_R$ tower:} \quad  &C(1; \lambda_{Z_R^{(n)}})  = 0 ~, \cr
\hbox{$\gamma$ tower:} \quad  &C'(1; \lambda_{\gamma^{(n)}})  = 0 ~, 
\label{Ztower2}
\end{align}
where 
\begin{align}
c_\phi = \frac{g_A}{\sqrt{g_A^2 + g_B^2}} ~, ~~ s_\phi = \frac{g_B}{\sqrt{g_A^2 + g_B^2}}  ~.
\label{angle1}
\end{align}
The spectra of the $Z$ and $\hat Z$  towers for $\sin \theta_H=0$ reduce to those determined by
$C'(1; \lambda_{Z^{(n)}})  = 0$ and $S(1; \lambda_{\hat Z^{(n)}})  = 0$, respectively.
The $Z$  tower has  zero mode $\lambda_{Z^{(0)}}$ for $\sin \theta_H=0$, whereas the $\hat Z$  tower
does not.  As in the case of $W$ and $\hat W$ towers,  the spectra of the $Z$ and $\hat Z$  towers alternate.
The lowest level ($Z^{(0)}$)  corresponds to $Z$ boson.   
A photon is always massless; $\lambda_{\gamma^{(0)}} = 0$.

The bare weak mixing angle $\theta_W^0$ is defined by
\begin{align}
\sin \theta_W^0 = \frac{s_\phi}{\sqrt{1 + s_\phi^2}} ~, ~~ 
\cos \theta_W^0 = \frac{1}{\sqrt{1 + s_\phi^2}}  ~.
\label{angle2}
\end{align}
It has been shown that in the case of $\theta_H^\min =0.1$ and $m_\KK =13\,$TeV,
for instance, $\sin^2 \theta_W^0 = 0.2305$ yields nearly the same phenomenology at low energies
as that of the SM with $\sin^2 \theta_W = 0.2312$.  In particular, it gives the forward-backward asymmetry 
$A_{FB} (e^- e^+ \go \mu^- \mu^+) = 0.01693$ at $\sqrt{s} = m_Z$.\cite{FCNC2020a, GUTinspired2020c}

$Z$ boson becomes massless in the $\theta_H=0$ and $\pi$ state.    As is seen from (\ref{Ztower1}), 
the wave function of $Z$ boson in the twisted gauge is nonvanishing in the $T^{3_R}$ and
$U(1)_\EM$ components for $\theta_H=\pi$.  
Note that $\tilde A_\mu = \sum_{a=1}^3 \{ \tilde A_\mu^{a_L} T^{a_L} + \tilde A_\mu^{a_R} T^{a_R}
+ \tilde A_\mu^{\hat a} T^{\hat a} \} +  \tilde A_\mu^{\hat 4} T^{\hat 4}$ and 
$g_A s_\phi = g_B c_\phi$.  
Inserting (\ref{Wtower1}) and  (\ref{Ztower1}) into  $g_A \tilde A_\mu + g_B Q_X B_\mu$, 
one finds that
gauge couplings of $W_\mu^{(0)}$, $Z_\mu^{(0)}$ and $A_\mu^{\gamma (0)}$ are given by
\begin{align}
&\underline{{\rm for~} \theta_H =0 :} \cr
\noalign{\kern 5pt}
&\frac{g_w}{\sqrt{2}} \, \Big\{ W_\mu^{(0)} (T^{1_L} + i T^{2_L}) + W_\mu^{(0)\dagger} (T^{1_L} - i T^{2_L}) \Big\} \cr
\noalign{\kern 5pt}
&\hskip 2.cm 
+ \frac{g_w}{\cos \theta_W^0}  \, Z_\mu^{(0)} \big( T^{3_L} - \sin^2 \theta_W^0 \, Q_\EM \big) 
+ e Q_\EM  \, A_\mu^{\gamma (0)}  \cr
\noalign{\kern 5pt}
&= g_w \Big\{ W_\mu^{1 (0)} T^{1_L} + W_\mu^{2 (0)} T^{2_L} + W_\mu^{3 (0)} T^{3_L} \Big\} 
+ g_w \tan \theta_W^0 B_\mu \, (T^{3_R} + Q_X) ~, \cr
\noalign{\kern 5pt}
&\underline{{\rm for~} \theta_H = \pi :} \cr
\noalign{\kern 5pt}
&\frac{g_w}{\sqrt{2}} \, \Big\{ W_\mu^{(0)} (T^{1_R} + i T^{2_R}) + W_\mu^{(0)\dagger} (T^{1_R} - i T^{2_R}) \Big\} \cr
\noalign{\kern 5pt}
&\hskip 2.cm 
+ \frac{g_w}{\cos \theta_W^0}  \, Z_\mu^{(0)} \big( T^{3_R} - \sin^2 \theta_W^0 \, Q_\EM \big) 
+ e Q_\EM  \, A_\mu^{\gamma (0)}  \cr
\noalign{\kern 5pt}
&= g_w \Big\{ W_\mu^{1 (0)} T^{1_R} + W_\mu^{2 (0)} T^{2_R} + W_\mu^{3 (0)} T^{3_R} \Big\} 
+ g_w \tan \theta_W^0 B_\mu \, (T^{3_L} + Q_X) ~, 
\label{gaugecoupling1}
\end{align}
where
\begin{align}
&g_w = \frac{g_A}{\sqrt{L}} ~, ~~ e = g_w \sin \theta_W^0 ~,~~ Q_\EM = T^{3_L} + T^{3_R} + Q_X  ~, \cr
\noalign{\kern 5pt}
&W_\mu^{(0)} = \frac{1}{\sqrt{2}} (W_\mu^{1 (0)} - i W_\mu^{2 (0)} ) , ~
\left[ \begin{matrix} Z_\mu^{(0)} \cr A_\mu^{\gamma (0)} \end{matrix} \right] = 
\left[ \begin{matrix} \cos \theta_W^0 & - \sin \theta_W^0 \cr  \sin \theta_W^0 & \cos \theta_W^0 \end{matrix} \right]
\left[ \begin{matrix} W_\mu^{3 (0)} \cr B_\mu  \end{matrix} \right]~.
\label{gaugecoupling1b}
\end{align}
In the $\theta_H = \pi$ state, there exists $SU(2)_R \times U(1)_{Y'}$ gauge symmetry in the twisted gauge
where $U(1)_{Y'}$ charge is given by $Y' = T^{3_L} + Q_X$.


Gauge couplings of quarks, leptons and dark fermions in the $\theta_H=\pi$ state are quite different from those
in the $\theta_H=0$ state.  For a fermion field $\Psi(x,z)$ it is most convenient to express its
KK expansion for $\check \Psi(x,z) = z^{-2} \Psi(x,z)$.

For up-type quarks in $\Psi^\alpha_{({\bf 3},{\bf 4})}$ in Table \ref{Tab:matterlist} 
the KK expansion is given, for the first generation pair $(u, u')$ for instance, by
\begin{align}
&\left[ \begin{matrix} \tilde{\check u} \cr \mynoalign \tilde{\check u}{}' \end{matrix} \right]
= \sqrt{k} \, \bigg\{ u^{(0)} +  \sum_{n=1}^\infty    u^{(n)} +    \sum_{n=1}^\infty  u^{\prime (n)} \bigg\} ~.
\label{waveUp0}
\end{align}
Wave functions are given in Appendix C.1.
The spectrum is determined  by
\begin{align}
\hbox{$u$ tower}: &\quad S_L S_R (1; \lambda_{u^{(n)}} , c_u) + \sin^2 \frac{\theta_H}{2} =  0 ~, \cr
\hbox{$u'$ tower}: &\quad S_L S_R (1; \lambda_{u^{\prime (n)}} , c_u) + \sin^2\frac{\theta_H}{2} = 0 ~,
\label{spectrumUp1}
\end{align}
where functions $S_{L/R}, C_{L/R}$ are given in (\ref{functionA2}).
We note $S_L S_R (1; \lambda , c) + \sin^2 \onehalf \theta_H = C_L C_R (1; \lambda , c) - \cos^2 \onehalf \theta_H$.
The lowest zero modes,  $\hat u^{(0)}_L(x)$ and  $\hat u^{(0)}_R(x)$ have chiral structure.
They are massless ($\lambda_{u^{(0)}} = 0$) for $\theta_H=0$.
Their wave functions behave differently from those of the $n\ge 1$ modes.   
The spectrum-determining equation for the $u$ tower for $c_u < 0$ reduces to
$S_R (1; \lambda_{u^{(n)}} , c_u) =0$ at $\theta_H=0$ and $C_L (1; \lambda_{u^{(n)}} , c_u) =0$ at $\theta_H=\pi$,
while for   the $u'$ tower it reduces to 
$S_L (1; \lambda_{u^{\prime (n)}} , c_u) =0$ at $\theta_H=0$ and $C_R (1; \lambda_{u^{\prime (n)}} , c_u) =0$ at $\theta_H=\pi$.
We note that  the spectrum $\{ \lambda_{u^{(n)}}, \lambda_{u^{\prime (n)}} \}$ at $\theta_H = \pi$ 
is different from that at $\theta_H = 0$.  
In particular $\lambda_{u^{(0)}} |_{\theta_H=0} = 0$ but $\lambda_{u^{(0)}} |_{\theta_H=\pi} > 0$.

Mass eigenstates of down-type quarks are more involved, the details of which have been given in Refs.\ 
\cite{GUTinspired2019a, FCNC2020a}.
Down-type quarks in $\Psi^\alpha_{({\bf 3},{\bf 4})}$ and  $\Psi^{\pm \alpha}_{({\bf 3},{\bf 1})}$ fields 
in Table \ref{Tab:matterlist}  mix with each other by brane interactions, which in the most general case induce
the CKM mass mixing matrix as well.  For the sake of simplicity we consider the case in which
brane interactions are diagonal in the generation space.
The spectrum for the first generation ($d$, $d'$ and $D^\pm$ towers) is determined by
\begin{align}
&\Big( S_L^Q S_R^Q +\sin^2\frac{\theta_H}{2} \Big)
 \big({\cal S}_{L1}^{D}{\cal S}_{R1}^{D}
 -{\cal S}_{L2}^{D}{\cal S}_{R2}^{D}\big) \cr
 \noalign{\kern 5pt}
 &\hskip 1.cm
+|\mu_1|^2 C_R^Q S_R^Q
 \left({\cal S}_{L1}^{D}{\cal C}_{L1}^{D} -{\cal S}_{L2}^{D}{\cal C}_{L2}^{D}\right)=0 
 \label{spectrumDown1}
\end{align}
where  $S_{L/R}^Q = S_{L/R}(1; \lambda, c_u)$, 
${\cal S}_{L j}^{D} = {\cal S}_{L j} (1; \lambda, c_{D_d}, \tilde m_{D_d})$, etc.
Functions ${\cal S}_{L/R j}$, $ {\cal C}_{L/R j}$ are given in (\ref{functionA3}).
$c_{D_\alpha}$ is the bulk mass parameter of $\Psi^{\pm \alpha}_{({\bf 3},{\bf 1})}$ field, and
$\tilde m_{D_\alpha} = m_{D_\alpha}/k$.
$\mu_1$ parametrizes  the strength of a brane interaction among $\Psi^{\alpha}_{({\bf 3},{\bf 4})}$, 
$\Psi^{\pm \alpha}_{({\bf 3},{\bf 1})}$ and $\Phi_S$, which is relevant to reproduce a mass of
each down-type quark.
There are four KK towers, ${\bf d} = (d, d', D^+, D^-)$.
Their KK expansion can be written  as 
\begin{align}
\left[ \begin{matrix} \tilde{\check d} \cr  \tilde{\check d}{}' \cr \tilde{\check D}{}^+ \cr  \tilde{\check D}{}^-  \end{matrix}  \right]
&= \sqrt{k} \bigg\{    \sum_{n=0}^\infty    d^{(n)} +    \sum_{n=1}^\infty  d^{\prime (n)} 
+ \sum_{n=1}^\infty  D^{+ (n)} + \sum_{n=1}^\infty  D^{- (n)}   \bigg\} .
\label{waveDown0}
\end{align}
Details of wave functions of all KK modes are given in Appendix C.1.
For $\theta_H=0$ there appears a massless mode $d^{(0)}$ with chiral wave functions, which is identified with a down quark.
For $\theta_H=\pi$ there is no zero mode.

$W$ and $Z$ couplings of quarks are easily found with the use of (\ref{gaugecoupling1}). 
We note that $\lambda_{u^{(n)}} = \lambda_{d^{(n)}}$ at $\theta_H=0$ and 
$\lambda_{u^{\prime (n)}} = \lambda_{d^{\prime (n)}}$ at $\theta_H=\pi$. 
Couplings with $W^{(0)}_\mu$, $Z^{(0)}_\mu$ and $A_\mu^{\gamma (0)}$are given by
\begin{align}
&\underline{{\rm for~} \theta_H =0 :} \cr
\noalign{\kern 5pt}
&\frac{g_w }{\sqrt{2}} \bigg\{ W^{(0)}_\mu  \, \Big( \bar{\hat u}_L^{(0)} \gamma^\mu \hat d_L^{(0)} 
+ \sum_{n=1}^\infty \bar{\hat u}^{(n)} \gamma^\mu \hat d^{(n)} \Big)
+W^{(0)\dagger}_\mu  \, \Big( \bar{\hat d}_L^{(0)} \gamma^\mu \hat u_L^{(0)} 
+ \sum_{n=1}^\infty \bar{\hat d}^{(n)} \gamma^\mu \hat u^{(n)} \Big) \bigg\}  \cr
\noalign{\kern 5pt}
&\hskip .5cm
+  \frac{g_w }{2 \cos \theta_W^0} Z^{(0)}_\mu \bigg\{ 
\big( \bar{\hat u}_L^{(0)} \gamma^\mu \hat u_L^{(0)} -
\bar{\hat d}_L^{(0)} \gamma^\mu \hat d_L^{(0)} \big)  
+ \sum_{n=1}^\infty \big( \bar{\hat u}^{(n)} \gamma^\mu \hat u^{(n)} -
\bar{\hat d}^{(n)} \gamma^\mu \hat d^{(n)} \big)  \bigg\} \cr
\noalign{\kern 5pt}
&\hskip 1.5cm
+\bigg\{  - g_w \frac{\sin^2 \theta_W^0 }{\cos \theta_W^0} Z^{(0)}_\mu   
+  e A_\mu^{\gamma (0)} \bigg\} J_{\EM}^\mu ~,  \cr
\noalign{\kern 5pt}
&\underline{{\rm for~} \theta_H =\pi :} \cr
\noalign{\kern 5pt}
& \frac{g_w }{\sqrt{2}} \bigg\{ W^{(0)}_\mu  \,  
\sum_{n=1}^\infty \bar{\hat u}^{\prime (n)} \gamma^\mu \hat d^{\prime (n)} 
+W^{(0)\dagger}_\mu  \, 
 \sum_{n=1}^\infty \bar{\hat d}^{\prime (n)} \gamma^\mu \hat u^{\prime (n)}  \bigg\}  \cr
\noalign{\kern 5pt}
&\hskip .5cm
+  \frac{g_w }{2 \cos \theta_W^0} Z^{(0)}_\mu 
 \sum_{n=1}^\infty \big( \bar{\hat u}^{\prime (n)} \gamma^\mu \hat u^{\prime (n)} -
\bar{\hat d}^{\prime (n)} \gamma^\mu \hat d^{\prime (n)} \big)   \cr
\noalign{\kern 5pt}
&\hskip 1.5cm
+\bigg\{  - g_w \frac{\sin^2 \theta_W^0 }{\cos \theta_W^0} Z^{(0)}_\mu   
+  e A_\mu^{\gamma (0)} \bigg\} J_{\EM}^\mu ~,
\label{gaugecouplingQuark1}
\end{align}
where 
\begin{align}
&J_{\EM}^\mu =
\sum_{n=0}^\infty \Big( \frac{2}{3} \bar{\hat u}^{(n)} \gamma^\mu \hat u^{(n)}
- \frac{1}{3} \bar{\hat d}^{(n)} \gamma^\mu \hat d^{(n)} \Big)
\cr
\noalign{\kern 5pt}
&\hskip .2cm
+ \sum_{n=1}^\infty \Big( \frac{2}{3} \bar{\hat u}^{\prime (n)} \gamma^\mu \hat u^{\prime (n)}
- \frac{1}{3} \bar{\hat d}^{\prime (n)} \gamma^\mu \hat d^{\prime (n)} 
- \frac{1}{3} \bar{\hat D}^{+(n)} \gamma^\mu \hat D^{+ (n)} 
- \frac{1}{3} \bar{\hat D}^{-(n)} \gamma^\mu \hat D^{- (n)} \Big).
\label{gaugecouplingQuark2}
\end{align}
At $\theta_H=0$ couplings of massless modes are chiral, but those of all massive modes 
are vector-like.
At $\theta_H=\pi$, $W^{(0)}_\mu$ couplings to $\bar{\hat u}_L^{(0)} \gamma^\mu \hat d_L^{(0)}$
and $\bar{\hat u}_R^{(0)} \gamma^\mu \hat d_R^{(0)}$ vanish, and
for $Z^{(0)}_\mu$ couplings of $\hat u^{(0)}$ and $\hat d^{(0)}$,  the $T^{3_R}$ part 
vanishes with only the $U(1)_\EM$ part surviving.
All modes are massive and their gauge couplings are vectorlike.
We also note that a top quark becomes very heavy at $\theta_H=\pi$, but all other quarks and leptons
remain light.  For $\theta_H^\min = 0.1, m_\KK=13\,$TeV, for instance, 
$m_t = 4.14\,$TeV and $m_c = 12.4\,$GeV  at $\theta_H=\pi$.

The gauge couplings in (\ref{gaugecouplingQuark1}) can be clearly and neatly understood from quantum 
numbers in the $\theta_H=0$ and $\pi$  phases as summarized in Table \ref{tab:charge1}.

\begin{table}[bth]
\renewcommand{\arraystretch}{1.2}
\begin{center}
\begin{tabular}{|c||c|c||c|c|}
\hline
&\multicolumn{2}{|c||}{$\theta_H = 0$ phase} &\multicolumn{2}{|c|}{$\theta_H = \pi$ phase} \\
\cline{2-5}
&$SU(2)_L$ & $U(1)_Y$ & $SU(2)_R$ & $U(1)_{Y'}$  \\
\hline
$\begin{matrix} u_L^{(0)} \cr d_L^{(0)} \end{matrix}$ & $\bm{2}$ & $\frac{1}{6}$  && \\
\cline{1-3}
$\begin{matrix} u_R^{(0)} \cr d_R^{(0)} \end{matrix}$ & $\bm{1}$ & $\begin{matrix} \frac{2}{3} \cr - \frac{1}{3} \end{matrix}$
&$\bm{1}$ & $\begin{matrix} \frac{2}{3} \cr - \frac{1}{3} \end{matrix}$ \\
\cline{1-3}
$\begin{matrix} u^{(n)} \cr d^{(n)} \end{matrix}$ & $\bm{2}$ & $\frac{1}{6}$  && \\
\hline
$\begin{matrix} u^{\prime (n)} \cr d^{\prime (n)} \end{matrix}$ & $\bm{1}$ & $\begin{matrix} \frac{2}{3} \cr - \frac{1}{3} \end{matrix}$
&$\bm{2}$ & $\frac{1}{6}$ \\
\hline
\end{tabular}
\caption{
Charge assignment of $u, u', d$ and $d'$ towers under  $SU(2)_L \times U(1)_{Y}$  in the $\theta_H = 0$ phase 
and $SU(2)_R \times U(1)_{Y'}$ in the $\theta_H = \pi$ phase.  The index $n$ runs as $n=1, 2, 3, \cdots $.  
Only the zero modes $u^{(0)}$ and $d^{(0)}$ in the  $\theta_H = 0$ phase are chiral.}
\label{tab:charge1}
\end{center}
\end{table}

Gauge couplings in the lepton sector are found in the same manner.  Details are given in Appendix C.2.
We note that the gauge couplings in the lepton sector,  given by (\ref{Leptoncoupling0}), 
can be summarized as in Table \ref{tab:charge1} 
for the quark sector.    One needs to replace $(u,d, u', d')$ by $(\nu_e, e, \nu_e', e')$, and 
$(\frac{1}{6}, \frac{2}{3}, - \frac{1}{3})$ in $U(1)_Y$ and $U(1)_{Y'}$ charges by $(- \frac{1}{2}, 0, -1)$ there.


Dark fermions in the spinor representation $\Psi_F^\beta$ (darkF fermions) are denoted, in the twisted gauge,
as 
\begin{align}
&\tilde{\check \Psi}_F (x,z) = \left[ \begin{matrix} \tilde{\check F}_1 \cr  \tilde{\check F}_2 \cr
\tilde{\check F}_1'  \cr \tilde{\check F}_2' \end{matrix} \right].
\label{darkFwave0}
\end{align}
The spectrum of $F$ and $F'$ towers,  $\{ \lambda_{F^{(n)}}, \lambda_{F^{\prime (n)}} \}$, 
is determined by
\begin{align}
&S_L S_R (1; \lambda_n , c_F) + \cos^2 \frac{\theta_H}{2} 
=  C_L C_R (1; \lambda_n , c_F) - \sin^2 \frac{\theta_H}{2}  =0 ~.
\label{darkFspectrum1}
\end{align}
There appear massless modes at $\theta_H=\pi$.
The spectrum-determining equation for the $F$ tower 
reduces to
$C_L (1; \lambda_{F^{(n)}} , c_F) =0$ at $\theta_H=0$ and $S_R (1; \lambda_{F^{(n)}} , c_F) =0$ 
at $\theta_H=\pi$, while for the $F'$ tower it reduces to 
$C_R (1; \lambda_{F^{\prime (n)}} , c_F)=0$ at $\theta_H=0$ and 
$S_L (1; \lambda_{F^{\prime (n)}} , c_F) =0$ at $\theta_H=\pi$.
For $c_F > 0$ zero modes appear in the $F'$ tower, and the KK expansion is given by
\begin{align}
&\left[ \begin{matrix} \tilde{\check F}_j \cr  \tilde{\check F}_j'  \end{matrix} \right] = 
\sqrt{k} \, \bigg\{ \sum_{n=1}^\infty F_j^{(n)} +  \sum_{n=0}^\infty F_j^{\prime (n)}  \bigg\} ~.
\label{darkFwave1a}
\end{align}
Wave functions of each mode are given in Appendix C.3.
For $c_F < 0$ zero modes appear in the $F$ tower, and the KK expansion is given by
\begin{align}
&\left[ \begin{matrix} \tilde{\check F}_j \cr  \tilde{\check F}_j'  \end{matrix} \right] = 
\sqrt{k} \, \bigg\{ \sum_{n=0}^\infty F_j^{(n)} +  \sum_{n=1}^\infty F_j^{\prime (n)}  \bigg\} ~.
\label{darkFwave1b}
\end{align}

We stress that massless modes appear at $\theta_H=\pi$ in the darkF sector.
Gauge couplings at $\theta_H=\pi$ are given, for $c_F > 0$,  by
\begin{align}
&\frac{g_w }{\sqrt{2}} \bigg\{  \bigg[ W^{(0)}_\mu  \,  \Big( 
\bar{\hat F}_{1R}^{\prime (0)} \gamma^\mu \hat F_{2R}^{\prime (0)} 
+ \sum_{n=1}^\infty  \bar{\hat F}_{1}^{\prime (n)} \gamma^\mu \hat F_{2}^{\prime (n)}  \Big) 
 + ~ \hbox{H.c.} ~ \bigg]  \bigg\} \cr
\noalign{\kern 5pt}
&\hskip 0.2cm
+ \frac{g_w }{2 \cos \theta_W^0} Z^{(0)}_\mu  \bigg\{
\bar{\hat F}_{1R}^{\prime (0)} \gamma^\mu \hat F_{1R}^{\prime (0)}
- \bar{\hat F}_{2R}^{\prime (0)} \gamma^\mu  \hat F_{2R}^{\prime (0)} 
+ \sum_{n=1}^\infty \big( \bar{\hat F}_{1}^{\prime (n)} \gamma^\mu \hat F_{1}^{\prime (n)}
- \bar{\hat F}_{2}^{\prime (n)} \gamma^\mu  \hat F_{2}^{\prime (n)} \big) \bigg\}   \cr
\noalign{\kern 5pt}
&\hskip 0.2cm
+ \bigg\{ - g_w \frac{\sin^2 \theta_W^0 }{\cos \theta_W^0} Z^{(0)}_\mu + e A_\mu^{\gamma (0)}  \bigg\}
\bigg\{ \sum_{n=1}^\infty \big( \bar{\hat F}_{1}^{(n)}, \bar{\hat F}_{2}^{(n)} \, \big) \gamma^\mu
 Q_\EM  \begin{pmatrix} \hat F_{1}^{(n)} \cr \hat F_{2}^{(n)} \end{pmatrix}
  \cr
\noalign{\kern 5pt}
&\hskip 5.8cm
+\sum_{n=0}^\infty 
\big( \bar{\hat F}_{1}^{\prime (n)}, \bar{\hat F}_{2}^{\prime (n)} \, \big) \gamma^\mu Q_\EM 
\begin{pmatrix} \hat F_{1}^{\prime (n)} \cr \hat F_{2}^{\prime (n)} \end{pmatrix} \bigg\} ~. 
\label{darkFcoupling1}
\end{align}
The gauge couplings in  (\ref{darkFcoupling1})  at $\theta_H=\pi$, as well as those   at $\theta_H=0$, 
can be  understood from quantum numbers in each phase
summarized in Table \ref{tab:charge2}.

\begin{table}[bth]
\renewcommand{\arraystretch}{1.2}
\begin{center}
\begin{tabular}{|c||c|c||c|c|}
\hline
&\multicolumn{2}{|c||}{$\theta_H = 0$ phase} &\multicolumn{2}{|c|}{$\theta_H = \pi$ phase} \\
\cline{2-5}
&$SU(2)_L$ & $U(1)_Y$ & $SU(2)_R$ & $U(1)_{Y'}$  \\
\hline
$\begin{matrix} F_1^{(n)} \cr F_2^{(n)} \end{matrix}$ & $\bm{2}$ & $\frac{1}{6}$  
& $\bm{1}$ & $\begin{matrix} \frac{2}{3} \cr - \frac{1}{3} \end{matrix}$ \\
\hline
$\begin{matrix} F_{1L}^{\prime (0)} \cr F_{2L}^{\prime (0)} \end{matrix}$ &&
& $\bm{1}$ &$\begin{matrix} \frac{2}{3} \cr - \frac{1}{3} \end{matrix}$\\
\cline{1-1}
\cline{4-5}
$\begin{matrix} F_{1R}^{\prime (0)} \cr F_{2R}^{\prime (0)} \end{matrix}$ & $\bm{1}$ 
& $\begin{matrix} \frac{2}{3} \cr - \frac{1}{3} \end{matrix}$
&$\bm{2}$     & $\frac{1}{6}$ \\
\cline{1-1}
\cline{4-5}
$\begin{matrix} F_{1}^{\prime (n)} \cr F_{2}^{\prime (n)} \end{matrix}$ &&& $\bm{2}$ & $\frac{1}{6}$   \\
\hline
\end{tabular}
\caption{
Charge assignment of $F_1, F_2, F_1'$ and $F_2'$ towers under  $SU(2)_L \times U(1)_{Y}$  
in the $\theta_H = 0$ phase and  $SU(2)_R \times U(1)_{Y'}$ in the $\theta_H = \pi$ phase
for $c_F > 0$.    The index $n$ runs as $n=1, 2, 3, \cdots $.  
Only the zero modes $F_1^{\prime (0)}$ and $F_2^{\prime (0)}$ in the  $\theta_H = \pi$ phase are chiral.}
\label{tab:charge2}
\end{center}
\end{table}

Formulas for $c_F < 0$ are obtained by replacing $\hat F_{j}^{\prime (0)}$ by $\hat F_{j}^{(0)}$.
One can flip the orbifold boundary conditions for $\Psi_F^\beta$, too.
By reversing the parity assignment for $\Psi_F^\beta$ in Table \ref{Tab:matterlist},
the role of left-handed and right-handed components are interchanged.

Formulas for $SU(3)_C$-singlet darkF$_\ell$ ($\Psi_{F_\ell}^\beta$) fields take the same form 
as for $SU(3)_C$-triplet darkF fields.
The KK expansions are the same.  The only change is in their $U(1)$ charges.
As in the lepton case, 
$(\frac{1}{6}, \frac{2}{3}, - \frac{1}{3})$ in $U(1)_Y$ and $U(1)_{Y'}$ charges 
should be replaced by $(- \frac{1}{2}, 0, -1)$.

It is observed that all modes of the darkF tower are massive and their gauge couplings are vectorlike
at $\theta_H = 0$, but there appear chiral massless modes at $\theta_H = \pi$.
In other words a theory with $\Psi_F^\beta$ but no $\Psi_{F_\ell}^\beta$ would become
anomalous at $\theta_H = \pi$.  
The anomaly cancellation is achieved with a set $(\Psi_F^\beta, \Psi_{F_\ell}^\beta)$
just as in the cancellation in the quark-lepton sector at $\theta_H = 0$.
We would like to stress that the appearance of a set $(\Psi_F^\beta, \Psi_{F_\ell}^\beta)$
is a natural consequence from the viewpoint of grand unification.  One set is contained
in the {\bf 32} representation in $SO(11)$ gauge-Higgs grand unification.\cite{HosotaniYamatsu2015, Furui2016}

Dark fermions in the vector representation $\Psi_V^{\pm \gamma}$ (darkV fermions) are,
as depicted in Fig.\ \ref{fig:spectrum1}, always massive and very heavy, and therefore
they do not affect the behavior of the model at the finite temperature $T \lesssim T^\LR_{\rm decay}$ very much. 
Their gauge couplings are summarized in Appendix C.4.  It is shown there that all couplings are vector-like.

\subsection{Chiral fermions, anomaly  and nontopological solitons}

In the above we have shown how the $\theta_H=0$ phase is smoothly connected to the $\theta_H=\pi$ phase
in GHU, and have clarified gauge couplings of fermions in each phase.
It is remarkable and intriguing that chiral fermions (quark and lepton multiplets) in the $\theta_H=0$ phase
are continuously transformed to vectorlike fermions in the $\theta_H=\pi$ phase.
The situation is reversed for dark fermions in the spinor representation (darkF fermions).
They are massive and vector-like  in the $\theta_H=0$ phase,  becoming chiral in the $\theta_H=\pi$ phase.
This fact immediately leads to an important question about anomalies.  What is the fate of anomalies  in the $\theta_H=0$ phase
when $\theta_H$ is continuously changed to  $\theta_H=\pi$?  More generally we need to  understand 
what kinds of anomalies arise in general $\theta_H$ states in GHU.

Another point of interest is the possibility of having $\theta_H=\pi$ solitons, nontopologoical solitons similar
to Fermi balls.\cite{Hong2020}
The lowest modes of darkF fermions are very heavy both in the $\theta_H=0$ phase and
in the  $\theta_H=\theta_H^{\rm min}$ state (the current universe) at $T=0$, whereas they become 
massless in the $\theta_H= \pi$ phase.  There can be a nontopological soliton such that in its  inside  $\theta_H=\pi$ and
massless darkF fermions are filled, but its outside is in the $\theta_H=0$ or $\theta_H^{\rm min}$ state.
Although the energy density of the  $\theta_H=\pi$ state is larger than that of the $\theta_H^{\rm min}$ state,
darkF fermions inside the ball cannot freely go outside as their masses in the $\theta_H^{\rm min}$ state are large.
The phase $\theta_H$ cannot changes from $\pi$ to $\theta_H^{\rm min}$ either for the same reason.
Pair annihilation processes are involved as well.  As shown in the preceding subsections, 
some of dark fermion pairs can annihilate to virtual gauge bosons, which subsequently annihilate to
quark-lepton pairs.  Gauge bosons need to tunnel out from the inside to the outside of the soliton
as quarks/leptons are heavy inside the soliton.
It would make more difficult for an object to decay, if the total darkF number is nonvanishing.
As a whole such an object can become stable.  Its size can be large. 
There may be important cosmological consequences of such solitons.

We would like to leave these intriguing questions for future investigation.
The existence of the $\theta_H=\pi$ state in GHU may have profound implications.

\section{Summary and discussions} 

In the present paper we have investigated the behavior of the GUT inspired $SO(5) \times U(1) \times SU(3)$ 
GHU model at finite temperature.  At zero temperature the EW symmetry
$SU(2)_L \times U(1)_Y$ is dynamically broken to $U(1)_\EM$ by the Hosotani mechanism. 
As the temperature is raised, the EW symmetry is restored around $T = 163\,$GeV.
We have  shown that the transition is of weakly first order just as in the SM in perturbation theory.
Although the EW symmetry breaking mechanism at $T=0$ is quite different from that in the SM,
the behavior at finite temperature  $T \lesssim  1\,$TeV  is almost the same as in the SM.
This is due to the fact that the particle spectrum at low energies is the same as in the SM.

As the temperature is increased further, a new feature emerges in GHU.
Above $T^\LR_{c1} \sim m_\KK$ the $\theta_H=0$ and $\theta_H=\pi$ states become
almost degenerate.   In the effective potential $V_\eff (\theta_H; T)$ these two states
are separated by a barrier so that domain structure will be formed as the universe expands
and the temperature drops to  $\sim T^\LR_{c1}$.  Eventually the $\theta_H=\pi$ state
becomes totally unstable for $T < T^\LR_{c2} \sim 2.3\,$TeV.
We have shown that the transition from the $\theta_H=\pi$ state to the $\theta_H=0$ state
rapidly takes place around $T= T^\LR_{\rm decay} \sim 2.6\,$TeV.

The $\theta_H=0$  and  $\theta_H=\pi$ states are characterized 
as the $SU(2)_L \times U(1)_Y$  and $SU(2)_R \times U(1)_{Y'}$ phases, respectively.  
$W$ boson, $Z$ boson and photon become gauge bosons of $SU(2)_R \times U(1)_{Y'}$
symmetry  in the $\theta_H=\pi$ state.
The transition from the $\theta_H=\pi$ state to the $\theta_H=0$ state is called as 
the left-right (LR) transition.
Gauge couplings of quarks, leptons  and dark fermions in the $SU(2)_R \times U(1)_{Y'}$ phase
differ from those in the $SU(2)_L \times U(1)_Y$ phase.
We showed, for instance, that quarks do not couple to $W$ boson and their couplings to $Z$ boson 
come  solely from the $U(1)_\EM$ part in the $\theta_H=\pi$ state.

In the history of the early Universe the LR  transition is a first-order phase transition, taking place
by tunneling through bubble nucleation.  GWs are produced  in this transition, 
the amount of which, however,  turns out  small.  
A GW signal from the LR transition is far below the reach of the sensitivity of LISA etc.

The 4D Higgs boson corresponds to the 4D fluctuation mode of the AB phase $\theta_H$ 
in the fifth dimension in GHU.  There is only one Higgs boson in the GHU model under investigation.
This single boson connects the $U(1)_\EM$ phase at $T=0$, the $SU(2)_L \times U(1)_Y$ phase
and the $SU(2)_R \times U(1)_{Y'}$ phase.
It would be of great interest to know physical consequences of the existence of 
the $SU(2)_R \times U(1)_{Y'}$ phase in the context of the history of the evolution of the early Universe.

\section*{Acknowledgements}

This work was supported in part by European Regional Development Fund-Project Engineering Applications 
of Microworld Physics (Grant No. CZ.02.1.01/0.0/0.0/16$\underline{~}$019/0000766) (Y.O.), 
by the National Natural Science Foundation of China (Grants No. 11775092, No. 11675061, 
No. 11521064, No. 11435003, and No. 11947213) (S. F.), by the International Postdoctoral Exchange 
Fellowship Program (S.F.), and by Japan Society for the Promotion of Science, Grants-in-Aid for Scientific 
Research, Grant No. JP19K03873 (Y.H.) and Grants No. JP18H05543 and No. JP19K23440 (N. Y.).

\vskip 1.cm

\appendix
\section{Basis functions} 
Wave functions of gauge fields and fermions are expressed in terms of the following basis functions.
For gauge fields we introduce
\begin{align}
 C(z; \lambda) &= \frac{\pi}{2} \lambda z z_L F_{1,0}(\lambda z, \lambda z_L) ~,  \cr
 S(z; \lambda) &= -\frac{\pi}{2} \lambda  z F_{1,1}(\lambda z, \lambda z_L) ~, \cr
 C^\prime (z; \lambda) &= \frac{\pi}{2} \lambda^2 z z_L F_{0,0}(\lambda z, \lambda z_L) ~,  \cr
S^\prime (z; \lambda) &= -\frac{\pi}{2} \lambda^2 z  F_{0,1}(\lambda z, \lambda z_L)~, \cr
\noalign{\kern 5pt}
 F_{\alpha, \beta}(u, v) &\equiv 
J_\alpha(u) Y_\beta(v) - Y_\alpha(u) J_\beta(v) ~,
\label{functionA1}
\end{align}
where $J_\alpha (u)$ and $Y_\alpha (u)$ are Bessel functions of  the first and second kind .
A relation $CS' - S C' = \lambda z$ holds.
For fermion fields with a bulk mass parameter $c$, we define 
\begin{align}
\begin{pmatrix} C_L \cr S_L \end{pmatrix} (z; \lambda,c)
&= \pm \frac{\pi}{2} \lambda \sqrt{z z_L} F_{c+\frac12, c\mp\frac12}(\lambda z, \lambda z_L) ~, \cr
\begin{pmatrix} C_R \cr S_R \end{pmatrix} (z; \lambda,c)
&= \mp \frac{\pi}{2} \lambda \sqrt{z z_L} F_{c- \frac12, c\pm\frac12}(\lambda z, \lambda z_L) ~.
\label{functionA2}
\end{align}
These functions satisfy $C_L C_R - S_L S_R=1$, $C_L  (z; \lambda, -c) = C_R  (z; \lambda, c)$
and $S_L  (z; \lambda, -c) = - S_R  (z; \lambda, c)$.  Also note that $S_{L/R} (z; 0, c) = 0$ and 
$C_{L/R} (z; 0, c) \not= 0$.
To treat down-type quarks and  dark fermions we also use
\begin{align}
{\cal C}_{L1}(z; \lambda, c, \tilde m) &= C_L(z; \lambda, c+\tilde{m})+C_L(z; \lambda, c-\tilde{m}) ~, \cr
{\cal C}_{L2}(z; \lambda, c, \tilde m) &= S_L(z; \lambda, c+\tilde{m})-S_L(z; \lambda, c-\tilde{m}) ~, \cr
{\cal S}_{L1}(z; \lambda, c, \tilde m) &= S_L(z; \lambda, c+\tilde{m})+S_L(z; \lambda,c-\tilde{m}) ~, \cr
{\cal S}_{L2}(z; \lambda, c, \tilde m) &= C_L(z; \lambda, c+\tilde{m})-C_L(z; \lambda, c-\tilde{m}) ~, \cr
{\cal C}_{R1}(z; \lambda, c, \tilde m) &= C_R(z; \lambda, c+\tilde{m})+C_R(z; \lambda, c-\tilde{m}) ~, \cr
{\cal C}_{R2}(z; \lambda, c, \tilde m) &= S_R(z; \lambda, c+\tilde{m})-S_R(z; \lambda,c-\tilde{m}) ~, \cr
{\cal S}_{R1}(z; \lambda, c, \tilde m) &= S_R(z; \lambda, c+\tilde{m})+S_R(z; \lambda, c-\tilde{m}) ~, \cr
{\cal S}_{R2}(z; \lambda, c, \tilde m) &= C_R(z; \lambda, c+\tilde{m})-C_R(z; \lambda, c-\tilde{m}) ~.
\label{functionA3}
\end{align}

\section{Gauge fields}

\def\myspace{\noalign{\kern 3pt}}

Wave functions of KK towers of gauge fields in the twisted gauge can be summarized in simple forms.
With $A_\mu^{a_{L/R}} = 2^{-1/2} (\onehalf \ep^{abc} A_\mu^{(bc)} \pm  A_\mu^{(a4)})$
and $A_\mu^{\hat a} = A_\mu^{a 4}$, 
a set $(A_\mu^{b_L}, A_\mu^{b_R}, A_\mu^{\hat b})$ ($b=1,2$)
forms  charged gauge field towers,  $W$, $\hat W$ and $W_R$  towers;
\begin{align}
\left[ \begin{matrix} \tilde A_\mu^{1_L} - i  \tilde A_\mu^{2_L} \cr \myspace
\tilde A_\mu^{1_R} - i  \tilde A_\mu^{2_R} \cr \myspace
\tilde A_\mu^{\hat 1} - i  \tilde A_\mu^{\hat 2} \end{matrix} \right]
&=  \left[ \begin{matrix} (1+c_H)  \mathring{W}_\mu \cr \myspace (1-c_H)  \mathring{W}_\mu \cr \myspace
- \sqrt{2} s_H  \mathring{W}^S_\mu  \end{matrix} \right]  + 
\left[ \begin{matrix} s_H (1+c_H) \mathring{\hat W}_\mu \cr  s_H (1-c_H) \mathring{\hat W}_\mu \cr
\sqrt{2} \,   \mathring{\hat W}^S_\mu \end{matrix} \right] +
\left[ \begin{matrix} (1-c_H) \mathring{W}_{R\mu}  \cr  \myspace
-(1+c_H) \mathring{W}_{R\mu}  \cr \myspace  0 \end{matrix} \right] ,   \cr
\noalign{\kern 10pt}
\left[ \begin{matrix} \mathring{W}_\mu(x,z) \cr \myspace \mathring{W}^S_\mu(x,z) \end{matrix} \right]
&= \sqrt{k} \sum_{n=0}^\infty W_\mu^{(n)} (x) 
\frac{1}{\sqrt{r_{W^{(n)}}}} 
\left[ \begin{matrix} C(z; \lambda_{W^{(n)}})  \cr \myspace \hat S(z; \lambda_{W^{(n)}}) \end{matrix} \right] ~,  \cr
\noalign{\kern 5pt}
\left[ \begin{matrix} \mathring{\hat W}_\mu(x,z) \cr \mathring{\hat W}^S_\mu(x,z) \end{matrix} \right]
&= \sqrt{k} \sum_{n=1}^\infty \hat W_\mu^{(n)} (x) 
\frac{1}{\sqrt{r_{\hat W^{(n)}}}} 
\left[ \begin{matrix} C(z; \lambda_{\hat W^{(n)}})  \cr \myspace \check S(z; \lambda_{\hat W^{(n)}}) 
\end{matrix} \right] ~,  \cr
\noalign{\kern 5pt}
\mathring{W}_{R\mu}(x,z) ~ &= \sqrt{k} \sum_{n=1}^\infty W_{R \mu}^{(n)} (x)  
\frac{1}{\sqrt{r_{W_R^{(n)}}}} \,  C(z; \lambda_{W_R^{(n)}}) ~,   \cr
\noalign{\kern 10pt}
r_{W^{(n)}} &= \int_1^{z_L} \frac{dz}{z} \big\{ (1+c_H^2) C(z; \lambda_{W^{(n)}})^2 
+ s_H^2 \hat S(z; \lambda_{W^{(n)}})^2 \big\} , \cr
r_{\hat W^{(n)}} &= \int_1^{z_L} \frac{dz}{z} \big\{ s_H^2 (1+c_H^2) C(z; \lambda_{\hat W^{(n)}})^2 
+ \check   S(z; \lambda_{\hat W^{(n)}})^2 \big\} , \cr
r_{W_R^{(n)}} &= \int_1^{z_L} \frac{dz}{z} \, (1+c_H^2) C(z; \lambda_{W_R^{(n)}})^2 , \cr
\noalign{\kern 5pt}
\hat S(z; \lambda) &= \frac{ C(1; \lambda)}{S(1; \lambda)} \,  S(z; \lambda), ~
\check S(z; \lambda) = \frac{ 2 C'(1; \lambda) C(1; \lambda)}{\lambda} \,  S(z; \lambda).
\label{Wtower1} 
\end{align}
Here $c_H = \cos \theta_H$, $s_H = \sin \theta_H$, and functions $C(z; \lambda), S(z; \lambda)$, etc.\ 
are defined in (\ref{functionA1}).
It is instructive to express the wave functions of $W$, $\hat W$ and $W_R$ towers  in the original gauge 
by making use of  (\ref{SO5generator2}).  For $\theta_H=\pi$, $A_\mu (x,z)$ in the original gauge is expanded as
\begin{align}
2 A_\mu (x,z) \Rightarrow
&\sum_{n=0}^\infty W_\mu^{(n)} (x) \frac{1}{\sqrt{r_{W^{(n)}}}} \, C(z; \lambda_{W^{(n)}})
\big\{ \big( 1 - \cos \theta(z) \big) (T^{1_L} + i T^{2_L}) \cr
&\hskip 1.cm 
+ \big( 1 +\cos \theta(z) \big) (T^{1_R} + i T^{2_R}) 
- \sin \theta (z) \sqrt{2} \, (T^{\hat 1} + i T^{\hat 2} ) \big\} \cr
\noalign{\kern 5pt}
+ & \sum_{n=1}^\infty \hat W_\mu^{(n)} (x) \frac{1}{\sqrt{r_{\hat W^{(n)}}}}  \, \check S(z; \lambda_{\hat W^{(n)}}) 
\big\{ \cos \theta (z) \sqrt{2}  \, (T^{\hat 1} + i T^{\hat 2} ) \cr
&\hskip 1.cm - \sin \theta (z) (T^{1_L} + i T^{2_L} - T^{1_R} - i T^{2_R}) \big\} \cr
\noalign{\kern 5pt}
+ &\sum_{n=1}^\infty W_{R\mu}^{(n)} (x) \frac{1}{\sqrt{r_{W_R^{(n)}}}} \, C(z; \lambda_{W_R^{(n)}})
\big\{ \big( 1 + \cos \theta(z) \big) (T^{1_L} + i T^{2_L}) \cr
&\hskip 1.cm 
+ \big( 1 -\cos \theta(z) \big) (T^{1_R} + i T^{2_R}) 
+ \sin \theta (z) \sqrt{2} \, (T^{\hat 1} + i T^{\hat 2} ) \big\} .
\label{Wtower3}
\end{align}
For $W(= W^{(0)})$ boson $\lambda_{W^{(0)}} =0$ and $C(z; \lambda_{W^{(0)}})/\sqrt{r_{W^{(0)}}} = 1/\sqrt{2kL}$.
It is seen that $W$ boson in the $\theta_H=\pi$ state is $SU(2)_L \,$-like at $z=1$, continuously changes 
in the group space $SO(5)$ in the bulk, and becomes $SU(2)_R \,$-like at $z=z_L$.  
$W_R$ tower, on the other hand, is $SU(2)_R \,$-like at $z=1$ and becomes $SU(2)_L \,$-like at $z=z_L$.
The brane interaction $\delta(y) \, (D_\mu \Phi_S)^\dagger D^\mu \Phi_S$ with the brane scalar $\Phi_S$
yields brane mass terms $\delta(y) \, \frac{1}{4} g_A^2 |w|^2 (A_\mu^{1_R} A^{1_R \mu} + A_\mu^{2_R} A^{2_R \mu} )$
after $ \Phi_S$ spontaneously develops vacuum-expectation-value $\la \Phi_{[1,2]} \ra = (0, w)^t \not= 0$.
Notice that this affects only $W_R$ tower in (\ref{Wtower3}).   $W$ boson becomes massless in the $\theta_H=\pi$ state.

Similarly in the sector of neutral gauge bosons, $(A_\mu^{3_L}, A_\mu^{3_R}, A_\mu^{\hat 3}, B_\mu)$, 
one finds in the twisted gauge that
\begin{align}
\left[ \begin{matrix} \tilde A_\mu^{3_L}  \cr \myspace \tilde A_\mu^{3_R}  \cr \myspace
\tilde A_\mu^{\hat 3} \cr \myspace B_\mu  \end{matrix} \right]
&=  \sqrt{\frac{1+s_\phi^2}{2}} 
\left[ \begin{matrix} (1+c_H)  \mathring{Z}_\mu  + s_H (1+c_H) \mathring{\hat Z}_\mu \cr 
(1-c_H)  \mathring{Z}_\mu + s_H (1-c_H) \mathring{\hat Z}_\mu\cr 
- \sqrt{2} s_H  \mathring{Z}^S_\mu  + \sqrt{2} \, (1+s_\phi^2)^{-1}  \mathring{\hat Z}^S_\mu  \cr  
0 \end{matrix} \right]  + 
\frac{1}{\sqrt{2}} \left[ \begin{matrix} (1-c_H) c_\phi   \cr  \myspace
-(1+c_H) c_\phi  \cr \myspace  0  \cr  2 c_H s_\phi \end{matrix} \right]\mathring{Z}_{R\mu}   \cr
\noalign{\kern 5pt}
&\hskip 4.cm
+ \frac{1}{\sqrt{1+s_\phi^2}} \left[ \begin{matrix} s_\phi \cr s_\phi \cr 0 \cr c_\phi  \end{matrix} \right]
\Big\{ \mathring{A}^\gamma_\mu - \sqrt{2} \, s_\phi \big( \mathring{Z}_\mu 
+ s_H \mathring{\hat Z}_\mu \big) \Big\}  ,   \cr
\noalign{\kern 10pt}
&\left[ \begin{matrix} \mathring{Z}_\mu(x,z) \cr \myspace \mathring{Z}^S_\mu(x,z) \end{matrix} \right]
= \sqrt{k} \sum_{n=0}^\infty Z_\mu^{(n)} (x) 
\frac{1}{\sqrt{r_{Z^{(n)}}}} 
\left[ \begin{matrix} C(z; \lambda_{Z^{(n)}})  \cr \myspace \hat S(z; \lambda_{Z^{(n)}}) \end{matrix} \right] ~,  \cr
\noalign{\kern 5pt}
&\left[ \begin{matrix} \mathring{\hat Z}_\mu(x,z) \cr \mathring{\hat Z}^S_\mu(x,z) \end{matrix} \right]
= \sqrt{k} \sum_{n=1}^\infty \hat Z_\mu^{(n)} (x) 
\frac{1}{\sqrt{r_{\hat Z^{(n)}}}} 
\left[ \begin{matrix} C(z; \lambda_{\hat Z^{(n)}})  \cr \myspace \check S(z; \lambda_{\hat Z^{(n)}}) 
\end{matrix} \right] ~,  \cr
\noalign{\kern 5pt}
&\mathring{Z}_{R\mu}(x,z) ~ = \sqrt{k} \sum_{n=1}^\infty Z_{R \mu}^{(n)} (x)  
\frac{1}{\sqrt{r_{Z_R^{(n)}}}} \,  C(z; \lambda_{Z_R^{(n)}}) ~,   \cr
\noalign{\kern 5pt}
&\mathring{A}^\gamma_\mu (x,z) ~ = \sqrt{k} \sum_{n=0}^\infty A_\mu^{\gamma (n)} (x)  
\frac{1}{\sqrt{r_{\gamma^{(n)}}}} \,  C(z; \lambda_{\gamma^{(n)}}) ~,  \cr
\noalign{\kern 10pt}
&r_{Z^{(n)}} = \int_1^{z_L} \frac{dz}{z} \big\{ [c_\phi^2 + (1+s_\phi^2) c_H^2] C(z; \lambda_{Z^{(n)}})^2 
+ (1+ s_\phi^2) s_H^2 \hat S(z; \lambda_{Z^{(n)}})^2 \big\} , \cr
&r_{\hat Z^{(n)}} = \int_1^{z_L} \frac{dz}{z} \big\{ 
s_H^2 [c_\phi^2 + (1+s_\phi^2) c_H^2] C(z; \lambda_{\hat Z^{(n)}})^2 
+ \frac{1}{1+s_\phi^2} \, \check   S(z; \lambda_{\hat Z^{(n)}})^2 \big\} , \cr
&r_{Z_R^{(n)}} = \int_1^{z_L} \frac{dz}{z} \, [c_\phi^2 + (1+s_\phi^2) c_H^2] C(z; \lambda_{Z_R^{(n)}})^2 , \cr
&r_{\gamma^{(n)}} = \int_1^{z_L} \frac{dz}{z} \,  C(z; \lambda_{Z_R^{(n)}})^2 ~.
\label{Ztower1} 
\end{align}

\section{Quarks, leptons and dark fermions}

Gauge couplings of quarks, leptons and dark fermions in the $\theta_H=\pi$ state are quite different from those
in the $\theta_H=0$ state.  For a fermion field $\Psi(x,z)$ it is most convenient to express its
KK expansion for $\check \Psi(x,z) = z^{-2} \Psi(x,z)$.

\subsection{Quark sector}

Wave functions of KK modes of up-type quarks in (\ref{waveUp0}) are given by
\begin{align}
&u^{(0)} =  \frac{\hat u^{(0)}_L(x)}{\sqrt{r_0}}  
\left[ \begin{matrix} \bar c_H C_L (z;  \lambda_{u^{(0)}}, c_u) \cr \mynoalign
i \bar s_H \hat S_L (z;  \lambda_{u^{(0)}}, c_u) \end{matrix} \right]
+  \frac{ \hat u^{(0)}_R(x) }{\sqrt{r_0}}  
 \left[ \begin{matrix} i \bar s_H \hat S_R (z;  \lambda_{u^{(0)}}, c_u)  \cr \mynoalign
\bar c_H   C_R (z;  \lambda_{u^{(0)}} c_u) \end{matrix} \right] , \cr 
\noalign{\kern 5pt}
& u^{(n)} =  \frac{ \hat u^{(n)}_L(x)}{\sqrt{r_n}} \, 
\left[ \begin{matrix} \bar c_H C_L (z;  \lambda_{u^{(n)}}, c_u) \cr \mynoalign
i \bar s_H \hat S_L (z;  \lambda_{u^{(n)}}, c_u) \end{matrix} \right] 
+ \frac{ \hat u^{(n)}_R(x)}{\sqrt{r_n}} \, 
 \left[ \begin{matrix} \bar c_H S_R (z;  \lambda_{u^{(n)}}, c_u)  \cr \mynoalign
i \bar s_H \hat C_R (z;  \lambda_{u^{(n)}}, c_u) \end{matrix} \right]  ,  \cr
\noalign{\kern 5pt}
& u^{\prime (n)} =  \frac{\hat u^{\prime (n)}_L(x)  }{\sqrt{r_n}} \,  
\left[ \begin{matrix}  i \bar s_H \hat C_L (z;  \lambda_{u^{\prime (n)}}, c_u) \cr \mynoalign
\bar c_H  S_L (z;  \lambda_{u^{\prime (n)}}, c_u) \end{matrix} \right] 
+ \frac{\hat u^{\prime (n)}_R(x)  }{\sqrt{r_n}} \,  
 \left[ \begin{matrix} i \bar s_H \hat S_R (z;  \lambda_{u^{\prime (n)}}, c_u)  \cr \mynoalign
\bar c_H C_R (z;  \lambda_{u^{\prime (n)}}, c_u) \end{matrix} \right] , 
\label{waveUp1}
\end{align}
where 
\begin{align}
&\bar c_H = \cos \onehalf \theta_H ~, ~~ \bar s_H = \sin \onehalf \theta_H ~, \cr
\noalign{\kern 5pt}
& \hat S_L (z;  \lambda, c) = N_L ( \lambda, c)  \, S_L (z;  \lambda, c) , ~
\hat C_R(z;  \lambda, c) = N_L ( \lambda, c)  \, C_R (z;  \lambda, c) , ~ \cr
\noalign{\kern 5pt}
&\hat S_R(z;  \lambda, c) = N_R ( \lambda, c) \, S_R (z;  \lambda, c) , ~
 \hat C_L (z;  \lambda, c) = N_R ( \lambda, c) \, C_L (z;  \lambda, c) , \cr
\noalign{\kern 5pt}
&N_L ( \lambda, c) =  \frac{C_L(1;  \lambda, c)}{S_L(1;  \lambda, c)} ~, ~~
N_R ( \lambda, c) =  \frac{C_R(1;  \lambda, c)}{S_R(1;  \lambda, c)} ~, 
\label{hatfunction1}
\end{align}
and a normalization factor $r_n$ should be understood in each term as
\begin{align}
&r_n = \int_1^{z_L} dz \, \Big\{ \big| f_n(z) \big|^2 + \big| g_n(z) \big|^2 \Big\} \quad
\hbox{in } \frac{1}{\sqrt{r_n}} \left[ \begin{matrix} f_n(z) \cr g_n(z) \end{matrix} \right] .
\label{normalization1}
\end{align}
Here ${\hat u}^{(n)}_L (x)$ and ${\hat u}^{\prime (n)}_L (x)$ 
(${\hat u}^{(n)}_R (x)$ and ${\hat u}^{\prime (n)}_R (x)$) are the left-handed (right-handed) components of 
4D fields ${\hat u}^{(n)} (x)$ and ${\hat u}^{\prime (n)} (x)$, respectively.

By making use of (\ref{spectrumUp1}) the expansion (\ref{waveUp1}) can be written as 
\begin{align}
&u^{(0)} =  \frac{\hat u^{(0)}_L(x)}{\sqrt{r_0}}  
\left[ \begin{matrix} \bar s_H C_L (z;  \lambda_{u^{(0)}}, c_u) \cr \mynoalign
- i \bar c_H \check S_L (z;  \lambda_{u^{(0)}}, c_u) \end{matrix} \right]
+  \frac{ \hat u^{(0)}_R(x) }{\sqrt{r_0}}  
 \left[ \begin{matrix} i \bar s_H  S_R (z;  \lambda_{u^{(0)}}, c_u)  \cr \mynoalign
\bar c_H   \check C_R (z;  \lambda_{u^{(0)}} c_u) \end{matrix} \right] , \cr 
\noalign{\kern 5pt}
& u^{(n)} =  \frac{ \hat u^{(n)}_L(x)}{\sqrt{r_n}} \, 
\left[ \begin{matrix} \bar s_H C_L (z;  \lambda_{u^{(n)}}, c_u) \cr \mynoalign
- i \bar c_H \check S_L (z;  \lambda_{u^{(n)}}, c_u) \end{matrix} \right] 
+ \frac{ \hat u^{(n)}_R(x)}{\sqrt{r_n}} \, 
 \left[ \begin{matrix} \bar s_H S_R (z;  \lambda_{u^{(n)}}, c_u)  \cr \mynoalign
- i \bar c_H \check C_R (z;  \lambda_{u^{(n)}}, c_u) \end{matrix} \right]  ,  \cr
\noalign{\kern 5pt}
& u^{\prime (n)} =  \frac{\hat u^{\prime (n)}_L(x)  }{\sqrt{r_n}} \,  
\left[ \begin{matrix}  - i \bar c_H \check C_L (z;  \lambda_{u^{\prime (n)}}, c_u) \cr \mynoalign
\bar s_H  S_L (z;  \lambda_{u^{\prime (n)}}, c_u) \end{matrix} \right] 
+ \frac{\hat u^{\prime (n)}_R(x)  }{\sqrt{r_n}} \,  
 \left[ \begin{matrix}-  i \bar c_H \check S_R (z;  \lambda_{u^{\prime (n)}}, c_u)  \cr \mynoalign
\bar s_H C_R (z;  \lambda_{u^{\prime (n)}}, c_u) \end{matrix} \right] , 
\label{waveUp2}
\end{align}
where
\begin{align}
& \check S_L (z;  \lambda, c) = N_R ( \lambda, c)^{-1}  \, S_L (z;  \lambda, c) , ~
\check C_R(z;  \lambda, c) = N_R ( \lambda, c)^{-1}  \, C_R (z;  \lambda, c) , ~ \cr
\noalign{\kern 5pt}
&\check S_R(z;  \lambda, c) = N_L ( \lambda, c)^{-1} \, S_R (z;  \lambda, c) , ~
\check C_L (z;  \lambda, c) = N_L ( \lambda, c)^{-1} \, C_L (z;  \lambda, c) .
\label{checkfunction1}
\end{align}
The expression in (\ref{waveUp2}) is more suitable at $\theta_H=\pi$ than that in (\ref{waveUp1}).
For the $u$ tower, for instance, $C_L (1;   \lambda_{u^{(n)}}, c_u)  =0$ so that $\hat S_L(z;  \lambda_{u^{(n)}}, c_u)$ 
and $\hat C_L(z;  \lambda_{u^{(n)}}, c_u)$ also vanish there.

The spectrum alternates as
$\lambda_{u^{(0)}}  <  \lambda_{u^{\prime (1)}} <  \lambda_{u^{(1)}} 
<  \lambda_{u^{\prime (2)}} <  \lambda_{u^{(2)}} < \cdots$.
Wave functions are given, up to normalization constants, by
\begin{align}
\hbox{for }\theta_H = 0, \quad
&\hat u^{(0)}_L: \left[ \begin{matrix} C_L(z; \lambda_{u^{(0)}}, c_u) \cr 0 \end{matrix} \right] , ~
\hat u^{(0)}_R: \left[ \begin{matrix} 0 \cr C_R(z; \lambda_{u^{(0)}}, c_u) \end{matrix} \right]  \cr
\noalign{\kern 5pt}
&\hat u^{(n)}_L: \left[ \begin{matrix} C_L(z; \lambda_{u^{(n)}}, c_u) \cr 0 \end{matrix} \right] , ~
\hat u^{(n)}_R: \left[ \begin{matrix} S_R(z; \lambda_{u^{(n)}}, c_u) \cr 0\end{matrix} \right] ~ (n \ge 1) \cr
\noalign{\kern 5pt}
&\hat u^{\prime (n)}_L: \left[ \begin{matrix} 0 \cr S_L(z; \lambda_{u^{\prime (n)}}, c_u) \end{matrix} \right] , ~
\hat u^{\prime(n)}_R: \left[ \begin{matrix} 0 \cr C_R(z; \lambda_{u^{\prime (n)}}, c_u) \end{matrix} \right] ~ (n \ge 1) 
\label{waveUp3}
\end{align}
and 
\begin{align}
\hbox{for }\theta_H = \pi, \quad
&\hat u^{(0)}_L: \left[ \begin{matrix} C_L(z; \lambda_{u^{(0)}}, c_u) \cr 0 \end{matrix} \right] , ~
\hat u^{(0)}_R: \left[ \begin{matrix}  i S_R(z; \lambda_{u^{(0)}}, c_u) \cr 0 \end{matrix} \right]  \cr
\noalign{\kern 5pt}
&\hat u^{(n)}_L: \left[ \begin{matrix} C_L(z; \lambda_{u^{(n)}}, c_u) \cr 0 \end{matrix} \right] , ~
\hat u^{(n)}_R: \left[ \begin{matrix} S_R(z; \lambda_{u^{(n)}}, c_u) \cr 0\end{matrix} \right] ~ (n \ge 1) \cr
\noalign{\kern 5pt}
&\hat u^{\prime (n)}_L: \left[ \begin{matrix} 0 \cr S_L(z; \lambda_{u^{\prime (n)}}, c_u) \end{matrix} \right] , ~
\hat u^{\prime(n)}_R: \left[ \begin{matrix} 0 \cr C_R(z; \lambda_{u^{\prime (n)}}, c_u) \end{matrix} \right] ~ (n \ge 1). 
\label{waveUp4}
\end{align}

Wave functions of KK modes of the four KK towers, ${\bf d} = (d, d', D^+, D^-)$, in (\ref{waveDown0}) are given by
\begin{align}
{\bf d}^{(n)} &= \hat {\bf d}^{(n)}_L(x)
\begin{pmatrix}  \alpha_{d} C_L(z; \lambda_{{\bf d}^{(n)}})\cr
\alpha_{d'}S_L(z; \lambda_{{\bf d}^{(n)}})\cr
a_{d}{\cal C}_{L2}(z; \lambda_{{\bf d}^{(n)}})+b_{d}{\cal C}_{L1}(z; \lambda_{{\bf d}^{(n)}}) \cr
a_{d}{\cal S}_{L1}(z; \lambda_{{\bf d}^{(n)}})+b_{d}{\cal S}_{L2}(z; \lambda_{{\bf d}^{(n)}})  \end{pmatrix}  \cr
\noalign{\kern 5pt}
&
+  \hat {\bf d}^{(n)}_R(x)
\begin{pmatrix}   \alpha_{d} S_R(z; \lambda_{{\bf d}^{(n)}}) \cr 
\alpha_{d'}C_R(z; \lambda_{{\bf d}^{(n)}}) \cr
a_{d}{\cal S}_{R2}(z; \lambda_{{\bf d}^{(n)}})+b_{d}{\cal S}_{R1}(z; \lambda_{{\bf d}^{(n)}}) \cr 
a_{d}{\cal C}_{R1}(z; \lambda_{{\bf d}^{(n)}})+b_{d}{\cal C}_{R2}(z; \lambda_{{\bf d}^{(n)}}) \cr 
\end{pmatrix}  
\label{waveDown1}
\end{align}
where $C_L(z; \lambda_{{\bf d}^{(n)}}) = C_L (z; \lambda_{{\bf d}^{(n)}}, c_u)$, 
${\cal C}_{Lj}(z; \lambda_{{\bf d}^{(n)}}) = {\cal C}_{Lj}(z;\lambda_{{\bf d}^{(n)}}, c_{D_d}, \tilde m_{D_d})$ 
and so on.
Coefficients $(\alpha_{d}, \alpha_{d'}, a_d, b_d)$ in each term satisfy 
\begin{align}
\begin{pmatrix}
\bar c_H  S_R^Q & -i \bar s_H C_R^Q & 0 & 0 \\
 -i \bar s_H C_L^Q & \bar c_H S_L^Q
 & \mu_1 {\cal C}_{L2}^{D} & \mu_1 {\cal C}_{L1}^{D}\\
-i\mu_1^* \bar s_H S_R^Q &
\mu_1^* \bar c_HC_R^Q & 
-{\cal S}_{R2}^{D} & -{\cal S}_{R1}^{D} \\
0 & 0 & {\cal S}_{L1}^{D} & {\cal S}_{L2}^{D}\\
\end{pmatrix}
\begin{pmatrix}
 \alpha_d\\
 \alpha_{d'}\\
 a_{d}\\
 b_{d}\\
\end{pmatrix}
= 0 
\label{waveDown2}
\end{align}
and an appropriate normalization condition, where $S_R^Q = S_R(1; \lambda)$, 
${\cal S}_{Lj}^D = {\cal S}_{Lj}(1; \lambda)$,  etc.

In the present paper it suffices to know wave functions  in the KK expansion for $\theta_H=0$ and $\pi$.
For  $\theta_H=0$ the condition matrix in (\ref{waveDown2}) becomes block-diagonal, and
(\ref{spectrumDown1}) and (\ref{waveDown2}) become
\begin{align}
&S_R^Q K^{0} = 0 , ~~ 
K^{0} = S_L^Q  \big({\cal S}_{L1}^{D}{\cal S}_{R1}^{D}  -{\cal S}_{L2}^{D}{\cal S}_{R2}^{D}\big) 
+ |\mu_1|^2 C_R^Q  
 \big({\cal S}_{L1}^{D}{\cal C}_{L1}^{D} -{\cal S}_{L2}^{D}{\cal C}_{L2}^{D}\big) , \cr
 \noalign{\kern 5pt}
&S_R^Q  \alpha_d = 0 ,~ 
\begin{pmatrix}
S_L^Q& \mu_1 {\cal C}_{L2}^{D} & \mu_1 {\cal C}_{L1}^{D}\\
\mu_1^*  C_R^Q &  -{\cal S}_{R2}^{D} & -{\cal S}_{R1}^{D} \\
 0 & {\cal S}_{L1}^{D} & {\cal S}_{L2}^{D}\\
\end{pmatrix}
\begin{pmatrix}  \alpha_{d'}\\  a_{d}\\  b_{d}\\   \end{pmatrix}
= 0 ~.
\label{spectrumDown2}
\end{align}
The KK tower specified by $S_R^Q =0$, $\{ \lambda_{d^{(n)}} \}$, contains a massless mode 
$\lambda_{d^{(0)}}=0$, which allows $\alpha_d \not= 0$ and a nontrivial left-handed mode.
$\lambda_{d^{(0)}}=0$ implies $K^0=0$ as well, and a nontrivial right-handed mode is 
contained in the $(d', D^+, D^-)$ components.
The spectrum determined by $S_R^Q \not= 0$ and $K^{0} = 0$ consists of three KK towers.
All modes are massive and $\alpha_{d} =0$.  Their wave functions are contained in the 
$(d', D^+, D^-)$ components.
One finds that
\begin{align}
&\underline{\theta_H = 0 : } \cr
&d^{(0)} =  \hat d^{(0)}_L(x)
\begin{pmatrix}  \alpha_{d} C_L(z)\cr  0\cr  0 \cr  0  \end{pmatrix} 
+ \hat d^{(0)}_R(x)
\begin{pmatrix}  0 \cr   \bar{\alpha}_{d'}C_R(z) \cr  \bar{a}_{d}{\cal S}_{R2}(z) \cr  
\bar{a}_{d}{\cal C}_{R1}(z) \cr  \end{pmatrix}, ~~  
\bar{a}_d = \mu_1^* \, \frac{C_R^Q}{{\cal S}_{R2}^D}  \, \bar{\alpha}_{d'} ~,  \cr
\noalign{\kern 5pt}
&d^{(n)} =  \hat d^{(n)}_L(x) \begin{pmatrix}  \alpha_{d} C_L(z)\cr 0\cr  0 \cr  0  \end{pmatrix} 
+ \hat d^{(n)}_R(x) \begin{pmatrix}  \alpha_{d} S_R(z)\cr 0\cr  0 \cr  0  \end{pmatrix} \quad (n \ge 1), \cr
\noalign{\kern 5pt}
&{\rm for ~} {\bf d} = d', D^+, D^- \cr
&{\bf d}^{(n)} = \hat {\bf d}^{(n)}_L(x)
\begin{pmatrix} 0 \cr
\alpha_{d'}S_L(z)\cr
a_{d}{\cal C}_{L2}(z)+b_{d}{\cal C}_{L1}(z) \cr
a_{d}{\cal S}_{L1}(z)+b_{d}{\cal S}_{L2}(z)  \end{pmatrix}  
+  \hat {\bf d}^{(n)}_R(x)
\begin{pmatrix}  0 \cr 
\alpha_{d'}C_R(z) \cr
a_{d}{\cal S}_{R2}(z)+b_{d}{\cal S}_{R1}(z) \cr 
a_{d}{\cal C}_{R1}(z)+b_{d}{\cal C}_{R2}(z) \cr    \end{pmatrix}  . 
\label{waveDown3}
\end{align}
For brevity $\lambda_{{\bf d}^{(n)}}$ in $C_L(z; \lambda_{{\bf d}^{(n)}})$, etc. has been suppressed 
in the above formulas.

For $\theta_H=\pi$  the condition matrix in (\ref{waveDown2}) becomes, in place of (\ref{spectrumDown2}), 
\begin{align}
&C_R^Q K^{\pi} = 0 , ~~ 
K^{\pi} = C_L^Q  \big({\cal S}_{L1}^{D}{\cal S}_{R1}^{D}  -{\cal S}_{L2}^{D}{\cal S}_{R2}^{D}\big) 
+ |\mu_1|^2 S_R^Q  
 \big({\cal S}_{L1}^{D}{\cal C}_{L1}^{D} -{\cal S}_{L2}^{D}{\cal C}_{L2}^{D}\big) , \cr
 \noalign{\kern 5pt}
&C_R^Q  \alpha_{d'} = 0 ,~ 
\begin{pmatrix}
-i C_L^Q& \mu_1 {\cal C}_{L2}^{D} & \mu_1 {\cal C}_{L1}^{D}\\
-i \mu_1^*  S_R^Q &  -{\cal S}_{R2}^{D} & -{\cal S}_{R1}^{D} \\
 0 & {\cal S}_{L1}^{D} & {\cal S}_{L2}^{D}\\
\end{pmatrix}
\begin{pmatrix}  \alpha_{d}\\  a_{d}\\  b_{d}\\   \end{pmatrix}
= 0 ~.
\label{spectrumDown3}
\end{align}
There is no massless mode.
The lowest mode $d^{(0)}$ is contained in one of the three KK towers determined by the conditions
$C_R^Q \not= 0$ and $K^{\pi} = 0$ for which $\alpha_{d'} =0$.  
(Recall that when $\mu_1=0$, the lowest mode satisfies
$C_L^Q = 0$ just as in the up-type quark spectrum.)
The spectrum of $d'$ tower is determined by $C_R^Q = 0$ for which $K^\pi \not= 0$ and 
$\alpha_d = a_d = b_d = 0$.  One finds that
\begin{align}
&\underline{\theta_H=\pi : } \cr
&d^{\prime (n)} =  \hat d^{\prime (n)}_L(x) \begin{pmatrix} 0 \cr   \alpha_{d'} S_L(z)\cr 0\cr  0 \end{pmatrix} 
+ \hat d^{\prime (n)}_R(x) \begin{pmatrix}  0 \cr \alpha_{d'} C_R(z)\cr 0\cr  0  \end{pmatrix} , \cr
\noalign{\kern 5pt}
&{\rm for ~} {\bf d} = d, D^+, D^- \cr
&{\bf d}^{(n)} = \hat {\bf d}^{(n)}_L(x)
\begin{pmatrix}  \alpha_{d}C_L(z)\cr 0 \cr
a_{d}{\cal C}_{L2}(z)+b_{d}{\cal C}_{L1}(z) \cr
a_{d}{\cal S}_{L1}(z)+b_{d}{\cal S}_{L2}(z)  \end{pmatrix}  
+  \hat {\bf d}^{(n)}_R(x)
\begin{pmatrix}   \alpha_{d}S_R(z) \cr 0 \cr
a_{d}{\cal S}_{R2}(z)+b_{d}{\cal S}_{R1}(z) \cr 
a_{d}{\cal C}_{R1}(z)+b_{d}{\cal C}_{R2}(z) \cr    \end{pmatrix}  .  
\label{waveDown4}
\end{align}

It is easy to find  $W$ and $Z$ couplings of quarks.
At $\theta_H=0$ the spectrum of both $u$ and $d$ towers   
is determined by $S_R(1; \lambda_n, c_u) =0$ so that $\lambda_{u^{(n)}} = \lambda_{d^{(n)}}$.  
Couplings with $W^{(0)}_\mu$, $Z^{(0)}_\mu$ and $A_\mu^{\gamma (0)}$  are obtained by inserting 
(\ref{waveUp3}) and (\ref{waveDown3}) into
\begin{align}
& \int_1^{z_L} dz \Bigg[ \big( \bar{\check u}, \bar{\check d} \, \big) \gamma^\mu \bigg\{ 
\frac{g_w }{\sqrt{2}} W^{(0)}_\mu (T^{1_L} + i T^{2_L}) 
+ \frac{g_w }{\sqrt{2}} W^{(0) \dagger}_\mu (T^{1_L} - i T^{2_L})   \cr
\noalign{\kern 5pt}
&\hskip 2.cm
+\frac{g_w }{\cos \theta_W^0} Z^{(0)}_\mu  (T^{3_L} - \sin^2 \theta_W^0 Q_\EM ) 
+ e A_\mu^{\gamma (0)} Q_\EM \bigg\}
\begin{pmatrix} \check u \cr \check d \end{pmatrix} \cr
\noalign{\kern 5pt}
&\hskip 1.5cm
+\bigg\{  - g_w \frac{\sin^2 \theta_W^0 }{\cos \theta_W^0} Z^{(0)}_\mu   
+  e A_\mu^{\gamma (0)} \bigg\}
 (  \bar{\check u}{}', \bar{\check d}{}', \bar{\check D}^+, \bar{\check D}^- ) \gamma^\mu
 Q_\EM 
\begin{pmatrix} \check u{}' \cr {\check d}{}' \cr {\check D}^+ \cr{\check D}^- \end{pmatrix} \Bigg],
\label{gaugecoupling2}
\end{align}
with  the fact that wave functions of gauge bosons are constant.
It leads to the expression in (\ref{gaugecouplingQuark1}) for $\theta_H=0$. 
The couplings of the zero modes  in (\ref{gaugecoupling2}) are the same as in the SM 
with $\theta_W^0$ replaced by $\theta_W$.

At $\theta_H=\pi$,  $SU(2)_R$ doublet components become relevant.  
Notice that the spectrum of both $u'$ and $d'$ towers is determined by $C_R(1, \lambda_n, c_u)=0$
and $\lambda_{u^{\prime (n)}} = \lambda_{d^{\prime (n)}}$.
Further $SU(2)_R$ components of the wave functions of $\hat u^{(n)}$ and $\hat d^{(n)}$ vanish.
It follows from  (\ref{gaugecoupling1}) gauge couplings are obtained  by inserting 
(\ref{waveUp4}) and (\ref{waveDown4}) into 
\begin{align}
& \int_1^{z_L} dz \Bigg[ \big( \bar{\tilde{\check u}}', \bar{\tilde{\check d}}'  \, \big) \gamma^\mu \bigg\{ 
\frac{g_w }{\sqrt{2}} W^{(0)}_\mu (T^{1_R} + i T^{2_R}) 
+ \frac{g_w }{\sqrt{2}} W^{(0) \dagger}_\mu (T^{1_R} - i T^{2_R})   \cr
\noalign{\kern 5pt}
&\hskip 2.cm
+\frac{g_w }{\cos \theta_W^0} Z^{(0)}_\mu  (T^{3_R} - \sin^2 \theta_W^0 Q_\EM ) 
+ e A_\mu^{\gamma (0)} Q_\EM \bigg\}
\begin{pmatrix} \tilde{\check u}' \cr \tilde{\check d}' \end{pmatrix} \cr
\noalign{\kern 5pt}
&\hskip 1.cm
+\bigg\{  - g_w \frac{\sin^2 \theta_W^0 }{\cos \theta_W^0} Z^{(0)}_\mu   
+  e A_\mu^{\gamma (0)} \bigg\}
 (  \bar{\tilde{\check u}}, \bar{\tilde{\check d}}, \bar{\tilde{\check D}}^+, \bar{\tilde{\check D}}^- )
  \gamma^\mu  Q_\EM 
\begin{pmatrix} \tilde{\check u} \cr \tilde{\check d} \cr \tilde{\check D}^+ \cr \tilde{\check D}^- \end{pmatrix} 
\Bigg]  . \label{gaugecoupling3}
\end{align}
It leads to the expression in (\ref{gaugecouplingQuark1}) for $\theta_H=\pi$.

\subsection{Lepton sector}

Charged lepton towers have the same form of KK expansions as up-type quark towers.  
For the first generation
\begin{align}
&\left[ \begin{matrix} \tilde{\check e} \cr \mynoalign \tilde{\check e}{}' \end{matrix} \right]
= \sqrt{k} \, \bigg\{ e^{(0)} +  \sum_{n=1}^\infty    e^{(n)} +    \sum_{n=1}^\infty  e^{\prime (n)} \bigg\} ~.
\label{waveElectron0}
\end{align}
The spectrum is determined by the same formula as (\ref{spectrumUp1}) where $c_u$ is replaced
by $c_e$. 
The expansions have the same form as (\ref{waveUp1}), (\ref{waveUp2}),  (\ref{waveUp3})
and (\ref{waveUp4}) where the replacement $(u, u') \go (e, e')$ should be made.

In the neutrino sector brane fermion $\chi$ satisfying the Majorana condition couples to $\nu$ and $\nu'$
through brane interactions.  In the two-component basis
\begin{align}
&\left[ \begin{matrix} \tilde{\check{\nu}}_L \cr \tilde{\check{\nu}}_L' \cr  \eta \end{matrix} \right]
 =\sqrt{k} \bigg\{  \sum_{n=0}^\infty ( \nu_{+L}^{(n)} +   \nu_{-L}^{(n)}  )  
 +  \sum_{n=1}^\infty (  \nu_{+L}^{\prime(n)} +   \nu_{-L}^{\prime (n)}  ) \bigg\} ~, \cr
\noalign{\kern 10pt}
&\left[ \begin{matrix}\tilde{\check{\nu}}_R \cr \tilde{\check{\nu}}_R' \cr \eta^c \end{matrix} \right]
 =\sqrt{k} \bigg\{  \sum_{n=0}^\infty ( \nu_{+R}^{(n)} +  \nu_{-R}^{(n)}  )  
 +  \sum_{n=1}^\infty ( \nu_{+R}^{\prime(n)} +   \nu_{-R}^{\prime (n)}  ) \bigg\} ~, 
\label{neutrinowave0}
\end{align}
where $\xi^c \equiv e^{i\delta_C} \sigma^2  \xi^*$ and $\chi = (\eta^c, \eta)$.
The spectrum is determined by
\begin{align}
(k\lambda \mp M )
\Big\{ S_L^L S_R^L +\sin^2\frac{\theta_H}{2} \Big\}
+ \frac{m_B^2}{k} S_R^L C_R^L=0 
\label{neutrinoSpectrum1}
\end{align}
for $\nu_\pm$ and $\nu_\pm'$ fields.
Here $S_R^L = S_R (1, \lambda_{\bm{\nu}^{(n)}}, c_e)$, etc., 
$M$ is a Majorana mass for $\chi$, and $m_B$ comes from a brane interaction
among $\nu'$, $\chi$ and $\Phi_S$.  
For $\theta_H \not= 0$ a tiny neutrino mass is generated
by gauge-Higgs  seesaw mechanism\cite{seesaw2017} similar to the inverse seesaw 
mechanism\cite{ Mohapatra1986}; 
$m_\nu \sim m_e^2 M/(2 |c_e|-1) m_B^2$  for $c_e < - \onehalf$.
Moderate values $M \sim 5\,$GeV, $m_B \sim 1\,$TeV yield $m_\nu \sim 1 \,$meV.

Wave functions of each mode in the expansion (\ref{neutrinowave0}) in  the neutrino sector
are given, with $\bm{\nu} = (\nu, \nu')$, 
\begin{align}
&\bm{\nu}_{\pm L}^{(n)}
 =\hat{\bm{\nu}}_{\pm L}^{(n)} (x)
 \begin{pmatrix} \alpha_\nu C_L (z, \lambda_{\bm{\nu}^{(n)}}, c_e) \cr 
  i\alpha_{\nu'} S_L(z, \lambda_{\bm{\nu}^{(n)}}, c_e)  \cr  i\alpha_\eta/\sqrt{k} \end{pmatrix} , ~
  \bm{\nu}_{\pm R}^{(n)}
 =\hat{\bm{\nu}}_{\pm R}^{(n)} (x)
 \begin{pmatrix} \alpha_\nu S_R (z, \lambda_{\bm{\nu}^{(n)}}, c_e) \cr 
  i\alpha_{\nu'} C_R(z, \lambda_{\bm{\nu}^{(n)}}, c_e)  \cr 
  \mp  i\alpha_\eta^* /\sqrt{k} \end{pmatrix} ,\cr
\noalign{\kern 10pt}
&\hat{\bm{\nu}}_{\pm L}^{(n)} (x)^c =  \pm \hat{\bm{\nu}}_{\pm R}^{(n)} (x) ~, 
\label{neutrinowave1}
\end{align}
where $\xi^c \equiv e^{i\delta_C} \sigma^2  \xi^*$.
Coefficients $(\alpha_\nu, \alpha_{\nu'}, \alpha_\eta)$ in each term satisfy
\begin{align}
\begin{pmatrix} 
\bar c_H S_R^L &\bar s_H C_R^L &0 \cr
-\bar s_H C_L^L &\bar c_H S_L^L& {m_B}/{k} \cr
m_B \bar s_H S_R^L &- m_B \bar c_H C_R^L 
& k\lambda \mp M \end{pmatrix}
\begin{pmatrix}   \alpha_\nu \cr  \alpha_{\nu'} \cr \alpha_\eta \end{pmatrix} =0 ~,
\label{neutrinowave2}
\end{align}
which leads to the spectrum-determining equation (\ref{neutrinoSpectrum1}).

For $\theta_H=0$  the condition (\ref{neutrinoSpectrum1}) reduces to
$S_R^L \big\{ (k\lambda \mp M ) S_L^L + k^{-1} {m_B^2}  C_R^L \big\} = 0$.
$\hat \nu$ tower with the spectrum determined by $S_R^L =0$ contains a massless left-handed 
neutrino ($\lambda_{\nu^{(0)}} = 0$).
\begin{align}
&\underline{\theta_H = 0} : \cr
\noalign{\kern 5pt}
&\nu_{+ L}^{(0)}
 =\hat{\nu}_{+ L}^{(0)} (x)  \begin{pmatrix} \alpha_\nu C_L (z) \cr 0  \cr 0\end{pmatrix} , ~
 \nu_{+ R}^{(0)} = 0 ~, \cr
\noalign{\kern 5pt}
 &\nu_{\pm L}^{(n)}
 =\hat{\nu}_{\pm L}^{(n)} (x) \begin{pmatrix} \alpha_\nu C_L (z) \cr   0  \cr 0 \end{pmatrix} , ~
 \nu_{\pm R}^{(n)}
 =\hat{\nu}_{\pm R}^{(n)} (x) \begin{pmatrix} \alpha_{\nu} S_R (z) \cr  0  \cr 0 \end{pmatrix} ~ (n \ge 1),  \cr
\noalign{\kern 5pt}
&\nu_{\pm L}^{\prime (n)}
 =\hat{\nu}_{\pm L}^{\prime (n)} (x) 
 \begin{pmatrix} 0 \cr i \alpha_{\nu'} S_L (z) \cr   i \alpha_\eta/\sqrt{k} \end{pmatrix} , ~
 \nu_{\pm R}^{\prime (n)}
 =\hat{\nu}_{\pm R}^{\prime (n)} (x) 
 \begin{pmatrix} 0 \cr i \alpha_{\nu'} C_R (z) \cr   \mp i \alpha_\eta^*/\sqrt{k}  \end{pmatrix} , \cr
\noalign{\kern 5pt}
 &\hskip 4.cm  S_L^L \alpha_{\nu'} + m_B k^{-1} \alpha_\eta = 0 ~,  ~(n \ge 1).
 \label{neutrinowave3}
\end{align}
For brevity $\lambda_{{\bm{\nu}^{(n)}}}$ and $c_e$ in $C_L (z, \lambda_{\nu^{(n)}}, c_e)$, etc.
have been suppressed.  There is no $\nu_{- }^{(0)}$ mode.
For $\theta_H=\pi$  the condition (\ref{neutrinoSpectrum1}) reduces to
$C_R^L \big\{ (k\lambda \mp M ) C_L^L + k^{-1} {m_B^2}  S_R^L \big\} = 0$.
There is no massless mode.  The spectrum of $\nu'$ tower is given by $C_R^L =0$.
\begin{align}
&\underline{\theta_H = \pi} : \cr
 &\nu_{\pm L}^{(n)}
 =\hat{\nu}_{\pm L}^{(n)} (x) 
 \begin{pmatrix} \alpha_\nu C_L (z) \cr   0  \cr i \alpha_\eta/\sqrt{k} \end{pmatrix} , ~
 \nu_{\pm R}^{(n)}
 =\hat{\nu}_{\pm R}^{(n)} (x) 
 \begin{pmatrix} \alpha_{\nu} S_R (z) \cr  0  \cr \mp i \alpha_\eta^*/\sqrt{k} \end{pmatrix} ~ ,  \cr
\noalign{\kern 5pt}
 &\hskip3.cm  - C_L^L \alpha_{\nu} + m_B k^{-1} \alpha_\eta = 0 ~,  \cr
\noalign{\kern 5pt}
&\nu_{\pm L}^{\prime (n)}
 =\hat{\nu}_{\pm L}^{\prime (n)} (x) 
 \begin{pmatrix} 0 \cr i \alpha_{\nu'} S_L (z) \cr  0 \end{pmatrix} , ~
\nu_{\pm R}^{\prime (n)}
 =\hat{\nu}_{\pm R}^{\prime (n)} (x) 
 \begin{pmatrix} 0 \cr i \alpha_{\nu'} C_R (z) \cr  0\end{pmatrix} , 
\label{neutrinowave4}
\end{align}
where $n \ge 0$ for $\nu_+$ and $n \ge 1$ for others.

We note that $\lambda_{\nu_+^{(n)}} = \lambda_{\nu_-^{(n)}}$ ($n \ge 1$) for $\theta_H=0$, and
that $\lambda_{\nu_+^{\prime (n)}} = \lambda_{\nu_-^{\prime (n)}}$ ($n \ge 1$) for $\theta_H=\pi$.
Couplings with $W^{(0)}_\mu$, $Z^{(0)}_\mu$ and $A_\mu^{\gamma (0)}$are given by
\begin{align}
&\underline{{\rm for~}\theta_H = 0} : \cr
\noalign{\kern 5pt}
&
\frac{g_w }{\sqrt{2}} \bigg\{ W^{(0)}_\mu  \, \Big( \bar{\hat \nu}_{eL}^{(0)} \gamma^\mu \hat e_L^{(0)} 
+ \sum_{n=1}^\infty \bar{\hat \nu}_e^{(n)} \gamma^\mu \hat e^{(n)} \Big)
+W^{(0)\dagger}_\mu  \, \Big( \bar{\hat e}_L^{(0)} \gamma^\mu \hat \nu_{eL}^{(0)} 
+ \sum_{n=1}^\infty \bar{\hat e}^{(n)} \gamma^\mu \hat \nu_e^{(n)} \Big) \bigg\}  \cr
\noalign{\kern 5pt}
&+  \frac{g_w }{2 \cos \theta_W^0} Z^{(0)}_\mu \bigg\{ 
\big( \bar{\hat \nu}_{eL}^{(0)} \gamma^\mu \hat \nu_{eL}^{(0)} -
\bar{\hat e}_L^{(0)} \gamma^\mu \hat e_L^{(0)} \big)  
+ \sum_{n=1}^\infty \big( \bar{\hat \nu}_e^{(n)} \gamma^\mu \hat \nu_e^{(n)} -
\bar{\hat e}^{(n)} \gamma^\mu \hat e^{(n)} \big)  \bigg\} \cr
\noalign{\kern 5pt}
&\hskip 1.5cm
+\bigg[  - g_w \frac{\sin^2 \theta_W^0 }{\cos \theta_W^0} Z^{(0)}_\mu   
+  e A_\mu^{\gamma (0)} \bigg] J_\EM^\mu ~ , \cr
\noalign{\kern 5pt}
&\underline{{\rm for~}\theta_H = \pi} : \cr
\noalign{\kern 5pt}
&
\frac{g_w }{\sqrt{2}} \bigg\{ W^{(0)}_\mu  \, 
 \sum_{n=1}^\infty \bar{\hat \nu}_e^{\prime (n)} \gamma^\mu \hat e^{\prime (n)}  
+W^{(0)\dagger}_\mu  \, 
\sum_{n=1}^\infty \bar{\hat e}^{\prime (n)} \gamma^\mu \hat \nu_e^{\prime (n)}  \bigg\}  \cr
\noalign{\kern 5pt}
&+  \frac{g_w }{2 \cos \theta_W^0} Z^{(0)}_\mu 
 \sum_{n=1}^\infty \big( \bar{\hat \nu}_e^{\prime (n)} \gamma^\mu \hat \nu_e^{\prime (n)} -
\bar{\hat e}^{\prime (n)} \gamma^\mu \hat e^{\prime (n)} \big)  
+\bigg[  - g_w \frac{\sin^2 \theta_W^0 }{\cos \theta_W^0} Z^{(0)}_\mu   
+  e A_\mu^{\gamma (0)} \bigg] J_\EM^\mu ~, \cr
\noalign{\kern 5pt}
&J_\EM^\mu = - \sum_{n=0}^\infty  \bar{\hat e}^{(n)} \gamma^\mu \hat e^{(n)}  
- \sum_{n=1}^\infty  \bar{\hat e}^{\prime (n)} \gamma^\mu \hat e^{\prime (n)} ~.
\label{Leptoncoupling0}
\end{align}

\subsection{Dark fermion (darkF $\Psi_F^\beta, \Psi_{F_\ell}^\beta$) sector}

Wave functions of each mode in the KK expansion of darkF fermions with $c_F >0$, (\ref{darkFwave1a}), 
are given by
\begin{align}
&F_j^{(n)} = \frac{\hat F_{jL}^{(n)} (x) }{\sqrt{r_n}}
 \left[ \begin{matrix} \bar c_H C_L (z;  \lambda_{F^{(n)}} , c_F) \cr \mynoalign
 i \bar s_H \check S_L  (z;  \lambda_{F^{(n)}} , c_F) \end{matrix} \right]  + 
 \frac{\hat F_{jR}^{(n)} (x) }{\sqrt{r_n}}
 \left[ \begin{matrix} \bar c_H S_R (z;  \lambda_{F^{(n)}} , c_F) \cr \mynoalign
 i \bar s_H \check C_R  (z;  \lambda_{F^{(n)}} , c_F) \end{matrix} \right] , \cr
\noalign{\kern 5pt}
&F_j^{\prime (n)} = \frac{\hat F_{jL}^{\prime (n)} (x) }{\sqrt{r_n}}
 \left[ \begin{matrix} i \bar s_H \check C_L (z;  \lambda_{F^{\prime (n)}} , c_F) \cr \mynoalign
\bar c_H S_L  (z;  \lambda_{F^{\prime (n)}} , c_F) \end{matrix} \right]  + 
 \frac{\hat F_{jR}^{\prime (n)} (x) }{\sqrt{r_n}}
 \left[ \begin{matrix} i \bar s_H \check S_R (z;  \lambda_{F^{\prime (n)}} , c_F) \cr \mynoalign
\bar c_H  C_R  (z;  \lambda_{F^{\prime (n)}} , c_F) \end{matrix} \right] .
\label{darkFwave2}
\end{align}
Making use of (\ref{darkFspectrum1}), one can write the wave functions in (\ref{darkFwave2}) as
\begin{align}
&F_j^{(n)} = \frac{\hat F_{jL}^{(n)} (x) }{\sqrt{r_n}}
 \left[ \begin{matrix} \bar s_H C_L (z;  \lambda_{F^{(n)}} , c_F) \cr \mynoalign
-  i \bar c_H \hat S_L  (z;  \lambda_{F^{(n)}} , c_F) \end{matrix} \right]  + 
 \frac{\hat F_{jR}^{(n)} (x) }{\sqrt{r_n}}
 \left[ \begin{matrix} \bar s_H S_R (z;  \lambda_{F^{(n)}} , c_F) \cr \mynoalign
 - i \bar c_H \hat C_R  (z;  \lambda_{F^{(n)}} , c_F) \end{matrix} \right] , \cr
\noalign{\kern 5pt}
&F_j^{\prime (0)} = \frac{\hat F_{jL}^{\prime (0)} (x) }{\sqrt{r_0}}
 \left[ \begin{matrix} i \bar s_H  C_L (z;  \lambda_{F^{\prime (0)}} , c_F) \cr \mynoalign
\bar c_H \hat S_L  (z;  \lambda_{F^{\prime (0)}} , c_F) \end{matrix} \right]  + 
 \frac{\hat F_{jR}^{\prime (0)} (x) }{\sqrt{r_0}}
 \left[ \begin{matrix} - i \bar c_H \hat S_R (z;  \lambda_{F^{\prime (0)}} , c_F) \cr \mynoalign
\bar s_H  C_R  (z;  \lambda_{F^{\prime (0)}} , c_F) \end{matrix} \right] , \cr
\noalign{\kern 5pt}
&F_j^{\prime (n)} = \frac{\hat F_{jL}^{\prime (n)} (x) }{\sqrt{r_n}}
 \left[ \begin{matrix} -i \bar c_H \hat C_L (z;  \lambda_{F^{\prime (n)}} , c_F) \cr \mynoalign
\bar s_H S_L  (z;  \lambda_{F^{\prime (n)}} , c_F) \end{matrix} \right]  + 
 \frac{\hat F_{jR}^{\prime (n)} (x) }{\sqrt{r_n}}
 \left[ \begin{matrix} -i \bar c_H \hat S_R (z;  \lambda_{F^{\prime (n)}} , c_F) \cr \mynoalign
\bar s_H  C_R  (z;  \lambda_{F^{\prime (n)}} , c_F) \end{matrix} \right] ,
\label{darkFwave3}
\end{align}
where $n \ge 1$.  The expression (\ref{darkFwave3})  is appropriate to use at $\theta_H = \pi$.  
In particular, notice that  $\lambda_{F^{\prime (0)}} = 0$,
$N_L(\lambda_{F^{\prime (0)}} , c_F)^{-1} =0$ and 
$\hat S_L (z; \lambda_{F^{\prime (0)}} , c_F)$ is finite there.

The spectrum alternates as
$\lambda_{F^{\prime (0)}}  <  \lambda_{F^{(1)}} <  \lambda_{F^{\prime (1)}} 
<  \lambda_{F^{(2)}} <  \lambda_{F^{\prime (2)}} < \cdots$.
Wave functions  are given, up to normalization constants, by
\begin{align}
\hbox{for }\theta_H = 0, \quad
&\hat F^{(n)}_{jL}: \left[ \begin{matrix} C_L(z; \lambda_{F^{(n)}}, c_F) \cr 0 \end{matrix} \right] , ~
\hat F^{(n)}_{jR}: \left[ \begin{matrix} S_R(z; \lambda_{F^{(n)}}, c_F) \cr 0\end{matrix} \right] ~ (n \ge 1) \cr
\noalign{\kern 5pt}
&\hat F^{\prime (n)}_{jL}: \left[ \begin{matrix} 0 \cr S_L(z; \lambda_{F^{\prime (n)}}, c_F) \end{matrix} \right] , ~
\hat F^{\prime(n)}_{jR}: \left[ \begin{matrix} 0 \cr C_R(z; \lambda_{F^{\prime (n)}}, c_F) \end{matrix} \right] ~ (n \ge 0) 
\label{darkFwave4}
\end{align}
and 
\begin{align}
\hbox{for }\theta_H = \pi, \quad
&\hat F^{(n)}_{jL}: \left[ \begin{matrix} C_L(z; \lambda_{F^{(n)}}, c_F) \cr 0 \end{matrix} \right] , ~
\hat F^{(n)}_{jR}: \left[ \begin{matrix} S_R(z; \lambda_{F^{(n)}}, c_F) \cr 0\end{matrix} \right] ~ (n \ge 1) \cr
\noalign{\kern 5pt}
&\hat F^{\prime (0)}_{jL}: \left[ \begin{matrix} i  C_L(z; \lambda_{F^{\prime (0)}}, c_F) \cr 0 \end{matrix} \right] , ~
\hat F^{\prime(0)}_{jR}: \left[ \begin{matrix} 0 \cr C_R(z; \lambda_{F^{\prime (0)}}, c_F) \end{matrix} \right] \cr
\noalign{\kern 5pt}
&\hat F^{\prime (n)}_{jL}: \left[ \begin{matrix} 0 \cr S_L(z; \lambda_{F^{\prime (n)}}, c_F) \end{matrix} \right] , ~
\hat F^{\prime(n)}_{jR}: \left[ \begin{matrix} 0 \cr C_R(z; \lambda_{F^{\prime (n)}}, c_F) \end{matrix} \right] ~ (n \ge 1). 
\label{darkFwave5}
\end{align}
Remember that $\lambda_{F^{\prime (0)}} |_{\theta_H=0} > 0$ 
but $\lambda_{F^{\prime (0)}} |_{\theta_H=\pi} = 0$.
Wave functions of the massless (zero) modes at $\theta_H = \pi$ are given, up to normalization factors, 
by
\begin{align}
&\tilde{\check \Psi}_F (x,z) 
\Rightarrow
\left[ \begin{matrix} \hat F_{1L}^{\prime (0)} (x)  \cr 
\hat F_{2L}^{\prime (0)} (x) \cr 0 \cr 0 \end{matrix} \right] i C_L(z; 0, c_F) , ~~
\left[ \begin{matrix} 0 \cr 0 \cr \hat F_{1R}^{\prime (0)} (x)  \cr 
\hat F_{2R}^{\prime (0)} (x)  \end{matrix} \right]  C_R(z; 0, c_F) ~. 
\label{darkFwave6}
\end{align}

For $c_F < 0$ zero modes appear in the $F$ tower, and the KK expansion is given by (\ref{darkFwave1b}).
The zero mode $F_j^{(0)}$ can be written as
\begin{align}
&F_j^{(0)} = \frac{\hat F_{jL}^{(0)} (x) }{\sqrt{r_0}}
 \left[ \begin{matrix} \bar s_H  C_L (z;  \lambda_{F^{(0)}} , c_F) \cr \mynoalign
- i \bar c_H \hat S_L  (z;  \lambda_{F^{(0)}} , c_F) \end{matrix} \right]  + 
 \frac{\hat F_{jR}^{(0)} (x) }{\sqrt{r_0}}
 \left[ \begin{matrix} \bar c_H \hat S_R (z;  \lambda_{F^{(0)}} , c_F) \cr \mynoalign
i \bar s_H  C_R  (z;  \lambda_{F^{ (0)}} , c_F) \end{matrix} \right] 
\label{darkFwave8}
\end{align}
and at $\theta_H=\pi$
\begin{align}
&\tilde{\check \Psi}_F (x,z) 
\Rightarrow
\left[ \begin{matrix} \hat F_{1L}^{(0)} (x)  \cr 
\hat F_{2L}^{(0)} (x) \cr 0 \cr 0 \end{matrix} \right] C_L(z; 0, c_F) , ~~
\left[ \begin{matrix} 0 \cr 0 \cr \hat F_{1R}^{(0)} (x)  \cr 
\hat F_{2R}^{(0)} (x)  \end{matrix} \right]  i C_R(z; 0, c_F) 
\label{darkFwave9}
\end{align}
which takes the same form as (\ref{darkFwave6}).

Gauge couplings at $\theta_H=\pi$ are given, for $c_F > 0$,  by
\begin{align}
& \int_1^{z_L} dz \Bigg[ \big( \bar{\tilde{\check F}}_1', \bar{\tilde{\check F}}_2'  \, \big) \gamma^\mu 
\bigg\{ 
\frac{g_w }{\sqrt{2}} W^{(0)}_\mu (T^{1_R} + i T^{2_R})  
+ \frac{g_w }{\sqrt{2}} W^{(0) \dagger}_\mu (T^{1_R} - i T^{2_R})   \cr
\noalign{\kern 5pt}
&\hskip 2.cm
+\frac{g_w }{\cos \theta_W^0} Z^{(0)}_\mu  (T^{3_R} - \sin^2 \theta_W^0 Q_\EM ) 
+ e A_\mu^{\gamma (0)} Q_\EM \bigg\}
\begin{pmatrix} \tilde{\check F}_1' \cr \tilde{\check F}_2' \end{pmatrix} \cr
\noalign{\kern 5pt}
&\hskip 2.cm
+\bigg\{  - g_w \frac{\sin^2 \theta_W^0 }{\cos \theta_W^0} Z^{(0)}_\mu   
+  e A_\mu^{\gamma (0)} \bigg\}
 (  \bar{\tilde{\check F}}_1, \bar{\tilde{\check F}}_2)
  \gamma^\mu  Q_\EM 
\begin{pmatrix} \tilde{\check F}_1 \cr \tilde{\check F}_2  \end{pmatrix}  \Bigg] ,   
\label{darkFcoupling0}
\end{align}
which leads to (\ref{darkFcoupling1}).

\subsection{Dark fermion (darkV $\Psi_V^\gamma$) sector}

DarkV field $\Psi_V^{\pm} = (\Psi_1^\pm, \cdots, \Psi_5^\pm)$ is in the representation $({\bf 1}, {\bf 5})_0$.  
Let us denote
\begin{align}
&\frac{1}{\sqrt{2}}
\begin{pmatrix} \Psi_4^\pm + i \Psi_3^\pm & \Psi_2^\pm + i \Psi_1^\pm \cr
- \Psi_2^\pm + i \Psi_1^\pm  & \Psi_4^\pm  - i \Psi_3^\pm \end{pmatrix} 
= \begin{pmatrix} N^\pm & -  E^{\prime \pm} \cr E^\pm & -  N^{\prime \pm} \end{pmatrix}, ~
\Psi_5^\pm = S^\pm ~.
\label{darkVname}
\end{align}
$N^\pm, N^{\prime \pm}, S^\pm$ are neutral ($Q_\EM =0$).   
$E^\pm$ and $E^{\prime \pm}$ have charges $Q_\EM = -1$ and $ +1$.
Under $SU(2)_L$ and $SU(2)_R$ rotations 
\begin{align}
&SU(2)_L: ~ \begin{pmatrix} N^\pm \cr E^\pm \end{pmatrix} , ~ 
\begin{pmatrix} E^{\prime \pm} \cr N^{\prime \pm} \end{pmatrix} \cr
\noalign{\kern 5pt}
&SU(2)_R: ~ \begin{pmatrix} N^{\prime \pm} \cr  E^\pm \end{pmatrix}, ~
\begin{pmatrix} E^{\prime \pm}  \cr N^\pm \end{pmatrix} 
\label{darkVsu2}
\end{align}
transform as  doublets.  $S^\pm$ fields are singlets.  We assume that bulk-mass parameters of
$\Psi_V^+$ and $\Psi_V^-$ are the same; $c_{V^+} = c_{V^-} = c_V$.

The mass spectrum is determined by
\begin{align}
&{\cal S}_{L1}^{V}{\cal S}_{R1}^{V}  -{\cal S}_{L2}^{V}{\cal S}_{R2}^{V} = 0 
\hskip 1.cm \hbox{for~}E^\pm , ~E^{\prime \pm} ,  \cr
\noalign{\kern 5pt}
&{\cal S}_{L1}^{V}{\cal S}_{R1}^{V}  -{\cal S}_{L2}^{V}{\cal S}_{R2}^{V} + 4 \sin^2 \theta_H = 0
\hskip 1.cm \hbox{for~} N^\pm, N^{\prime \pm}, S^\pm , 
\label{DarkVspectrum1}
\end{align}
where ${\cal S}_{L1}^{V} = {\cal S}_{L1} (1; \lambda, c_V, \tilde m_V)$, $\tilde m_V=m_V/k$, etc.
There are no zero modes.  As seen in Fig.\ \ref{fig:spectrum1}, the lowest mode has a mass $\sim 3\,$TeV.
For $\theta_H=0$ and $\pi$, the neutral and charged sectors become degenerate in masses.

$E^+$ ($E^{\prime +}$) fields couple to $E^-$ ($E^{\prime -}$) fields by Dirac mass terms.
The KK expansions are given, with the notation $\E= (E, E')$ and $\N = (N, N')$,  by 
\begin{align}
&\left[ \begin{matrix} \tilde{\check \E}{}^+ \cr  \tilde{\check \E}{}^- \end{matrix} \right] 
= \sqrt{k} \,  \sum_{n=1}^\infty \E^{(n)} ~, \cr
\noalign{\kern 5pt}
&\left[ \begin{matrix} \tilde{\check \N}{}^+ \cr  \tilde{\check \N}{}^- \end{matrix} \right] = 
\sqrt{k} \,  \sum_{n=1}^\infty \N^{(n)} ~, ~~
\left[ \begin{matrix} \tilde{\check S}{}^+ \cr  \tilde{\check S}{}^- \end{matrix} \right] = 
\sqrt{k} \,  \sum_{n=1}^\infty S^{(n)}  ~.
\label{darkVwave0}
\end{align}
Wave functions of each KK mode in (\ref{darkVwave0}) are given by 
\begin{align}
\E^{(n)} &=\frac{\hat \E_{L}^{(n)} (x) }{\sqrt{r_n}} 
\left[ \begin{matrix} a \, {\cal C}_{L2} (z;  \lambda_{E^{(n)}}) 
+ b \, {\cal C}_{L1} (z;  \lambda_{E^{(n)}})  \cr \mynoalign
a \, {\cal S}_{L1} (z;  \lambda_{E^{(n)}} ) 
+ b \, {\cal S}_{L2} (z;  \lambda_{E^{(n)}} ) \end{matrix} \right]  \cr
\noalign{\kern 5pt}
&+
 \frac{\hat \E_{R}^{(n)} (x) }{\sqrt{r_n}}
 \left[ \begin{matrix} a \, {\cal S}_{R2} (z;  \lambda_{E^{(n)}} ) 
+ b \, {\cal S}_{R1} (z;  \lambda_{E^{(n)}} )  \cr \mynoalign
a \, {\cal C}_{R1} (z;  \lambda_{E^{(n)}}) 
+ b \, {\cal C}_{R2} (z;  \lambda_{E^{(n)}} ) \end{matrix} \right], \cr
\noalign{\kern 5pt}
&{\cal C}_{Lj} (z;  \lambda_{E^{(n)}})  = {\cal C}_{Lj} (z;  \lambda_{E^{(n)}} , c_V, \tilde m_V), ~ {\rm etc.}
\label{darkVwave1}
\end{align}
where ${\cal S}_{L1}^V a + {\cal S}_{L2}^V b = {\cal S}_{R2}^V a + {\cal S}_{R1}^V b  = 0$ in each term.
Note that $ \lambda_{E^{(n)}} =  \lambda_{E^{\prime (n)}}$.

In the neutral sector all fields $N^\pm, N^{\prime \pm}, S^\pm$ couple with each other for general 
$\theta_H$, but they split into pairs $(N^+, N^-)$, $(N^{\prime +}, N^{\prime -})$ and  $(S^+, S^-)$ 
for  $\theta_H =0$ and $\pi$. The spectrum becomes degenerate; 
$\lambda_{N^{(n)}} = \lambda_{N^{\prime (n)}} =\lambda_{S^{(n)}} = \lambda_{E^{(n)}}$.
Wave functions are given by 
\begin{align}
\N^{(n)} = &\frac{\hat \N_{L}^{(n)} (x) }{\sqrt{r_n}} 
\left[ \begin{matrix} a \, {\cal C}_{L2} (z;  \lambda_{N^{(n)}} ) 
+ b \, {\cal C}_{L1} (z;  \lambda_{N^{(n)}} )  \cr \mynoalign
a \, {\cal S}_{L1} (z;  \lambda_{N^{(n)}} ) 
+ b \, {\cal S}_{L2} (z;  \lambda_{N^{(n)}} ) \end{matrix} \right]  \cr
\noalign{\kern 5pt}
+ &
 \frac{\hat \N_{R}^{(n)} (x) }{\sqrt{r_n}}
 \left[ \begin{matrix} a \, {\cal S}_{R2} (z;  \lambda_{N^{(n)}} ) 
+ b \, {\cal S}_{R1} (z;  \lambda_{N^{(n)}} )  \cr \mynoalign
a \, {\cal C}_{R1} (z;  \lambda_{N^{(n)}}) 
+ b \, {\cal C}_{R2} (z;  \lambda_{N^{(n)}}) \end{matrix} \right], \cr
\noalign{\kern 5pt}
S^{(n)} = &\frac{\hat S_{L}^{(n)} (x) }{\sqrt{r_n}} 
\left[ \begin{matrix} a \, {\cal S}_{L1} (z;  \lambda_{S^{(n)}}) 
+ b \, {\cal S}_{L2} (z;  \lambda_{S^{(n)}})  \cr \mynoalign
a \, {\cal C}_{L2} (z;  \lambda_{S^{(n)}} ) 
+ b \, {\cal C}_{L1} (z;  \lambda_{S^{(n)}}) \end{matrix} \right]  \cr
\noalign{\kern 5pt}
+ &
 \frac{\hat S_{R}^{(n)} (x) }{\sqrt{r_n}}
 \left[ \begin{matrix} a \, {\cal C}_{R1} (z;  \lambda_{S^{(n)}}) 
+ b \, {\cal C}_{R2} (z;  \lambda_{S^{(n)}} )  \cr \mynoalign
a \, {\cal S}_{R2} (z;  \lambda_{S^{(n)}} ) 
+ b \, {\cal S}_{R1} (z;  \lambda_{S^{(n)}} ) \end{matrix} \right] ,
\label{darkVwave2}
\end{align}
where ${\cal S}_{L1}^V a + {\cal S}_{L2}^V b = {\cal S}_{R2}^V a + {\cal S}_{R1}^V b  = 0$ in each term.

Gauge couplings of darkV fields at $\theta_H=0$ and $\pi$ are easily found.
We note that the mass spectrum and wave functions of $\E^{(n)}$  are the same as those of $\N^{(n)}$.
$S^\pm$ fields do not couple to $W$, $Z$, and $A^\gamma$ fields.  One finds that 
\begin{align}
&\underline{{\rm for~}\theta_H = 0} : \cr
\noalign{\kern 5pt}
&
\frac{g_w }{\sqrt{2}} \bigg\{ W^{(0)}_\mu  \, 
\sum_{n=1}^\infty \Big( \bar{\hat N}^{(n)} \gamma^\mu \hat E^{(n)} + 
 \bar{\hat E}^{\prime (n)} \gamma^\mu \hat N^{\prime (n)}\Big)  + ~\hbox{H.c.} ~ \bigg\}  \cr
\noalign{\kern 5pt}
&+  \frac{g_w }{2 \cos \theta_W^0} Z^{(0)}_\mu 
\sum_{n=1}^\infty \Big( \bar{\hat N}^{(n)} \gamma^\mu \hat N^{(n)} -
\bar{\hat E}^{(n)} \gamma^\mu \hat E^{(n)} 
+  \bar{\hat E}^{\prime (n)} \gamma^\mu \hat E^{\prime (n)} 
- \bar{\hat N}^{\prime (n)} \gamma^\mu \hat N^{\prime (n)} \Big)   \cr
\noalign{\kern 5pt}
&+\bigg\{  - g_w \frac{\sin^2 \theta_W^0 }{\cos \theta_W^0} Z^{(0)}_\mu   
+  e A_\mu^{\gamma (0)} \bigg\} J_\EM^\mu ~,  \cr
\noalign{\kern 5pt}
&\underline{{\rm for~}\theta_H = \pi} : \cr
\noalign{\kern 5pt}
&
\frac{g_w }{\sqrt{2}} \bigg\{ W^{(0)}_\mu  \, 
\sum_{n=1}^\infty \Big( \bar{\hat N}^{\prime (n)} \gamma^\mu \hat E^{(n)} + 
 \bar{\hat E}^{\prime (n)} \gamma^\mu \hat N^{(n)}\Big)  + ~\hbox{H.c.} ~ \bigg\}  \cr
\noalign{\kern 5pt}
&+  \frac{g_w }{2 \cos \theta_W^0} Z^{(0)}_\mu 
\sum_{n=1}^\infty \Big( \bar{\hat N}^{\prime (n)} \gamma^\mu \hat N^{\prime (n)} -
\bar{\hat E}^{(n)} \gamma^\mu \hat E^{(n)} 
+  \bar{\hat E}^{\prime (n)} \gamma^\mu \hat E^{\prime (n)} 
- \bar{\hat N}^{(n)} \gamma^\mu \hat N^{ (n)} \Big)   \cr
\noalign{\kern 5pt}
&+\bigg\{  - g_w \frac{\sin^2 \theta_W^0 }{\cos \theta_W^0} Z^{(0)}_\mu   
+  e A_\mu^{\gamma (0)} \bigg\}  J_\EM^\mu ~, \cr
\noalign{\kern 5pt}
&J_\EM^\mu = \sum_{n=1}^\infty  \Big( - \bar{\hat E}^{(n)} \gamma^\mu \hat E^{(n)}  
+  \bar{\hat E}^{\prime (n)} \gamma^\mu \hat E^{\prime (n)}  \Big) ~.
\label{darkVcoupling}
\end{align}
All couplings are vectorlike.

\vskip .3cm

\def\jnl#1#2#3#4{{#1}{\bf #2},  #3 (#4)}

\def\Zphys{{\em Z.\ Phys.} }
\def\jssc{{\em J.\ Solid State Chem.\ }}
\def\jpsJ{{\em J.\ Phys.\ Soc.\ Japan }}
\def\ptps{{\em Prog.\ Theoret.\ Phys.\ Suppl.\ }}
\def\PTP{{\em Prog.\ Theoret.\ Phys.\  }}
\def\PTEP{{\em Prog.\ Theoret.\ Exp.\  Phys.\  }}
\def\JMP{{\em J. Math.\ Phys.} }
\def\NPB{{\em Nucl.\ Phys.} B}
\def\NP{{\em Nucl.\ Phys.} }
\def\PLB{{\it Phys.\ Lett.} B}
\def\PL{{\em Phys.\ Lett.} }
\def\PRL{\em Phys.\ Rev.\ Lett. }
\def\PRB{{\em Phys.\ Rev.} B}
\def\PRD{{\em Phys.\ Rev.} D}
\def\PRe{{\em Phys.\ Rep.} }
\def\AP{{\em Ann.\ Phys.\ (N.Y.)} }
\def\RMP{{\em Rev.\ Mod.\ Phys.} }
\def\ZPC{{\em Z.\ Phys.} C}
\def\SCI{\em Science}
\def\CMP{\em Comm.\ Math.\ Phys. }
\def\MPLA{{\em Mod.\ Phys.\ Lett.} A}
\def\IJMPA{{\em Int.\ J.\ Mod.\ Phys.} A}
\def\IJMPB{{\em Int.\ J.\ Mod.\ Phys.} B}
\def\EPJC{{\em Eur.\ Phys.\ J.} C}
\def\PR{{\em Phys.\ Rev.} }
\def\JHEP{{\em JHEP} }
\def\JCAP{{\em JCAP} }
\def\cmp{{\em Com.\ Math.\ Phys.}}
\def\JPA{{\em J.\  Phys.} A}
\def\JPG{{\em J.\  Phys.} G}
\def\NJP{{\em New.\ J.\  Phys.} }
\def\CQG{\em Class.\ Quant.\ Grav. }
\def\ATMP{{\em Adv.\ Theoret.\ Math.\ Phys.} }
\def\ibid{{\em ibid.} }
\def\ChP{{\em Chin.Phys.}C}


\renewenvironment{thebibliography}[1]
         {\begin{list}{[$\,$\arabic{enumi}$\,$]}  
         {\usecounter{enumi}\setlength{\parsep}{0pt}
          \setlength{\itemsep}{0pt}  \renewcommand{\baselinestretch}{1.2}
          \settowidth
         {\labelwidth}{#1 ~ ~}\sloppy}}{\end{list}}

\leftline{\Large \bf References}


\end{document}